# TransResAI: A Compound AI System for Coastal Transportation Resilience

Qingwen Pu[a], Kun Xie[a,*], and Chenyu Yan[a]

[a] *Transportation Informatics Lab, Department of Civil and Environmental Engineering, Old Dominion University, Norfolk, VA 23529, United States*
*Emails: qpu001@odu.edu (Pu), kxie@odu.edu (Xie), and cyan004@odu.edu (Yan)*
[*] *Corresponding Author*

## Abstract

Coastal flooding increasingly threatens transportation infrastructure, yet the analytical tools needed for resilience management remain difficult for many non-specialist practitioners to use. This study presents TransResAI, a compound AI system that supports analysis of flood-aware transportation resilience via natural-language interactions. The system integrates a locally deployable Large Language Model (LLM) with modules for task decomposition, secure code generation, geospatial analysis, retrieval-augmented generation, and interactive map rendering. TransResAI links MATSim flood-scenario simulation outputs, OpenStreetMap-derived flood-risk networks, equity-focused demographic indicators, and regional documents in Hampton Roads, Virginia. A structured user study with domain experts demonstrated that TransResAI reduced task completion time by 80–88% relative to conventional GIS workflows, compressing analytical tasks from a mean of 197.1 seconds to 29.7 seconds and visualization tasks from 364.0 seconds to 46.1 seconds, while maintaining mean accuracy of 4.60/5.00 and task completion rates exceeding 94%. These findings demonstrate that compound AI architectures bridge the gap between general-purpose language models and specialized domain knowledge, as well as the quantitative rigor required for infrastructure resilience, providing transportation agencies and communities with faster, more accessible analytical tools for decision-making under growing climate uncertainty.

## 1. Introduction

Coastal flooding imposes cascading disruptions on the transportation networks that underpin mobility, economic activity, and emergency response for hundreds of millions of people in low-lying urban regions worldwide (Pregnolato *et al.* 2017). In the United States, over 40% of the population resides in coastal counties particularly exposed to these risks (Nelson *et al.* 2022). Despite the urgency of this challenge, effective resilience management requires coordinating across multiple analytical demands, including identifying vulnerable infrastructure, planning evacuation routes, and targeting support for at-risk populations, tasks that current tools are ill-equipped to address at the speed and accessibility that practitioners need (Prashar *et al.* 2023, Aziz *et al.* 2024).

Existing approaches for transportation resilience management demonstrate significant limitations in accessibility and integration (Zhou *et al.* 2026). Traffic simulation methods require specialized expertise in configuring parameters and interpreting outputs, creating barriers for practitioners lacking simulation training (Zhang *et al.* 2024a). GIS-based assessments reduce resilience analysis to spatial overlay operations without capturing behavioral responses or network-wide cascading impacts (Arvidsson *et al.* 2023). Machine learning approaches capture statistical associations but fail to incorporate causal mechanisms underlying flood impacts on infrastructure and travel behavior (Museru *et al.* 2025). Consequently, current methods not only fail to provide decision-makers with accessible, comprehensive insights across diverse scenarios, but also concentrate analytical capacity among technically trained specialists, systematically excluding the community stakeholders and under-resourced agencies most exposed to flood risk (Habib *et al.* 2023, Chabot and Bertrand 2025).

Large Language Models (LLMs) offer a paradigm shift from specialized technical tools to natural language-based intelligent systems. LLMs can interpret plain language queries, decompose complex tasks into structured subtasks, and integrate information from heterogeneous sources while maintaining conversational interfaces accessible to non-technical users (Pu *et al.* 2026). Recent advances in Retrieval-Augmented Generation (RAG) enhance LLMs' ability to ground responses in external knowledge bases, reducing hallucination risks and incorporating domain-specific information (Zhao *et al.* 2024). Nevertheless, translating these general capabilities into operationally reliable tools for infrastructure analysis requires bridging fundamental gaps in domain knowledge, spatial reasoning, and institutional deployment constraints (Wang 2026).

Despite this potential, direct application of LLMs to transportation resilience management remains challenging (Xu *et al.* 2025). General-purpose LLMs lack specialized domain knowledge to interpret transportation metrics or understand causal relationships between flood depths, capacity reductions, and network redistribution (Zhang *et al.* 2024a). Resilience analysis requires joint reasoning across spatial networks, temporal dynamics, and demographic attributes—modalities that existing LLM architectures process as text sequences without explicit spatial or temporal mechanisms (Zhang *et al.* 2024b). Practical deployment also faces constraints related to computational costs and the need for transparent, auditable decision-making processes that agencies can validate and trust (Subramanian 2024). Addressing these challenges requires a compound AI framework integrating LLMs with domain-specific tools and structured reasoning modules (Chen *et al.* 2025a, Wu *et al.* 2025).

This study introduces TransResAI, an LLM-powered compound AI system providing a scalable framework for assessing infrastructure resilience to flooding through natural language interaction. TransResAI integrates an LLM with specialized modules for query parsing, task decomposition, multi-source information retrieval, and response generation. The framework is validated using comprehensive data from the Hampton Roads region, which exhibits extensive coastal exposure, diverse socioeconomic vulnerability patterns, and infrastructure configurations representative of mid-sized U.S. coastal cities. The main contributions include:

- A locally deployable methodological framework that transforms natural language queries into domain-specific analytical workflows, providing better data control and privacy than API-based alternatives.
- A spatially aware analytical capability that processes geospatial layers and generates visualizations, overcoming a key limitation of mainstream LLMs in handling spatial data.
- A multi-source integration mechanism that consolidates fragmented and heterogeneous data into a unified retrieval framework.

## 2. Literature Review

### 2.1. Transportation Resilience Assessment Methods

Transportation resilience assessment has evolved from static network analysis toward increasingly sophisticated behavioral and predictive approaches, yet remains constrained by accessibility barriers that limit practical deployment (Kammouh and Chahrour 2025). Early topology-based methods employed graph-theoretic measures such as betweenness centrality and link criticality to identify structurally important road segments (Jana *et al.* 2023). While computationally efficient, these approaches treat infrastructure as abstract networks without accounting for traffic demand, traveler behavior, or the differential impacts of partial versus complete link failures (Ldchn *et al.* 2025). The inability to quantify functional performance degradation motivated the development of dynamic modeling frameworks (Fang *et al.* 2022, Pan *et al.* 2022).

Agent-based traffic simulation platforms emerged as the dominant paradigm for capturing behavioral responses to infrastructure disruptions (Huang *et al.* 2022, He *et al.* 2026). MATSim and similar microsimulation tools represent individual travelers making utility-maximizing decisions about routes, departure times, and mode choices (Zhuge *et al.* 2021, Choi *et al.* 2025). These models successfully predict complex phenomena including spatial trip redistribution, temporal demand shifting, and cascading congestion effects following flood events (Li *et al.* 2025, He *et al.* 2026). However, their adoption in planning practice has been limited by the specialized expertise required for model calibration and the computational intensity that restricts scenario exploration (Zhuge *et al.* 2021). A metropolitan-scale simulation can require hours to days on high-performance computing infrastructure, making interactive analysis infeasible (Jiang *et al.* 2024).

The limitations of simulation-based approaches motivated parallel development of Geographic Information Systems (GIS) workflows for resilience assessment (Kammouh and Chahrour 2025). GIS platforms enable spatial overlay analysis combining flood inundation maps with infrastructure and demographic layers, providing accessible visualization capabilities for planning agencies (Nwogu *et al.* 2025). Yet GIS-based assessments reduce resilience analysis to identifying which facilities fall within flood zones, without quantifying functional impacts on traffic flow or network performance (Arvidsson *et al.* 2023, Lin *et al.* 2024). Standard workflows require repeating multi-step geoprocessing sequences for each scenario variation, creating time costs that limit comprehensive exploration of uncertainty ranges in flood projections (Singha *et al.* 2024). Data-driven approaches offer a complementary paradigm, using machine learning to predict traffic disruptions from historical flood observations (Yuan *et al.* 2023, Walker *et al.* 2024). These models achieve computational efficiency but sacrifice interpretability and struggle to generalize beyond their training distributions, limiting utility for scenarios that deviate from historical patterns (Lee *et al.* 2024, Museru *et al.* 2025).

The field thus faces a persistent accessibility-capability tradeoff that is particularly consequential for flood resilience planning, where practitioners must rapidly evaluate multiple inundation scenarios, integrate demographic vulnerability data, and communicate findings to non-technical stakeholders under time pressure (Chabot and Bertrand 2025, Konev *et al.* 2025). None of these paradigms integrate quantitative traffic analysis with policy knowledge synthesis or provide natural language interfaces for scenario analysis, a gap that recent advances in LLMs for urban infrastructure domains have begun to address (Nie *et al.* 2025, Ying *et al.* 2025). This motivates the development of a multi-module architecture that democratizes resilience planning by enabling non-specialist staff to perform sophisticated analyses via natural language, thereby bypassing the traditional barriers of GIS and programming proficiency (Bhanye 2025, Wang and Wu 2025).

### 2.2. LLM-Based Infrastructure Decision Support

Recent advances in LLMs have demonstrated transformative potential for domain-specific decision support, yet direct application to infrastructure domains faces fundamental challenges distinct from general-purpose question answering (Ling *et al.* 2025). Engineering decision support demands capabilities that general-purpose LLMs do not provide, including the ability to execute numerical computations, orchestrate domain-specific tools, and reason jointly across heterogeneous data modalities (Nguyen and Kittur 2025, Wu *et al.* 2025). Addressing these demands requires moving beyond general-purpose language model capabilities toward compound AI architectures specifically designed for infrastructure analytical workflows (Zaharia *et al.* 2024, Kandogan *et al.* 2025).

Retrieval-Augmented Generation (RAG) architectures have been proposed to ground LLM responses in domain-specific knowledge bases, reducing hallucination risks while enabling access to external information (Gao *et al.* 2023). Recent implementations in transportation network management and urban infrastructure planning demonstrate RAG's effectiveness for document retrieval and regulatory guidance synthesis (Lander *et al.* 2024, Zhu *et al.* 2024). However, these systems retrieve relevant text passages through semantic search without executing numerical computations or orchestrating external analytical tools, limiting their utility for engineering tasks requiring numerical simulation, spatial analysis, or quantitative scenario comparison (Liang *et al.* 2025, Sharma 2025).

Transportation resilience analysis presents additional challenges through its reliance on heterogeneous data modalities (Yang *et al.* 2025). Assessments require joint reasoning across spatial networks, temporal dynamics, demographic attributes, and policy constraints (Chen and Chopra 2025, Rouhana and Jawad 2025). Yet existing LLM architectures process all such inputs as flat text sequences, lacking explicit spatial or temporal reasoning mechanisms (Gurnee and Tegmark 2023, Mohsin *et al.* 2025). While recent progress in multimodal LLMs has advanced vision-language integration, geospatial and tabular data integration remains a distinct and largely open challenge, as text-based representations of structured spatial data discard relational geometry and numerical distributions critical for infrastructure performance assessment (Ji *et al.* 2025, Zhang *et al.* 2025).

A further and institutionally consequential limitation concerns data privacy and governance (Chen *et al.* 2025b). Existing LLM-based systems depend on cloud-hosted API services, including GPT-5, Claude, and Gemini, which require transmitting query content to external servers (Gupta and Shrivastava 2025). In government and infrastructure planning contexts, analytical workflows routinely involve sensitive demographic records, socioeconomic vulnerability indices, and location-specific operational data whose transmission to third-party platforms raises significant privacy and data governance concerns (Alrahamneh *et al.* 2025). Many transportation agencies handle data subject to strict confidentiality requirements, rendering cloud-dependent LLM deployments institutionally infeasible for most government transportation authorities (Yan *et al.* 2025). This motivates the development of locally deployed compound AI systems capable of executing the full analytical pipeline within agency-controlled computational environments, without reliance on external API services.

### 2.3. Research gaps

The preceding review identifies three specific gaps that existing frameworks have not resolved. First, RAG-based systems remain confined to text passage retrieval and cannot execute numerical computations, invoke simulation tools, or perform quantitative scenario comparison, all capabilities essential for engineering-grade resilience assessment (Liang *et al.* 2025, Wampler *et al.* 2025). Second, existing LLM-based frameworks reduce geospatial networks, tabular simulation outputs, and demographic vulnerability indicators to flat text sequences, discarding the relational geometry and numerical structure critical for infrastructure reasoning (Pierdicca *et al.* 2025, Truong *et al.* 2025). Third, the data governance constraints of government transportation agencies render cloud-dependent LLM systems operationally infeasible, a barrier that current compound AI architectures have not addressed (Alrahamneh *et al.* 2025). Collectively, these gaps define an unoccupied design space: a locally deployable compound AI system capable of translating natural language queries into quantitative resilience workflows while operating within agency-controlled computational environments.

## 3. Data

This study centers on the Hampton Roads region in southeastern Virginia, USA, a major coastal metropolitan area located at the confluence of the Chesapeake Bay and the Atlantic Ocean. The region is subject to land subsidence and accelerating sea-level rise driven by global warming, making this low-lying area particularly vulnerable to flooding and storm surges. Its dense transportation network is highly fragmented by waterways and heavily relies on bridges and tunnels, many of which serve as critical links carrying high traffic volume. The combination of these physical and infrastructural vulnerabilities makes the Hampton Roads region a particularly relevant and representative case for developing and evaluating transportation resilience assessment frameworks. The case study area comprises ten independent cities (Chesapeake, Franklin, Hampton, Newport News, Norfolk, Poquoson, Portsmouth, Suffolk, Virginia Beach, Williamsburg) and five counties (Gloucester, Isle of Wight, James City, Southampton, York).

This study constructed a multi-source external knowledge base to support LLM's RAG system. The knowledge base aggregated three core datasets: a flood risk network, traffic simulation outputs under baseline and flooding scenarios, and demographic information. Figure 1 provides an overview of the end-to-end data processing workflow for generating and integrating these datasets.

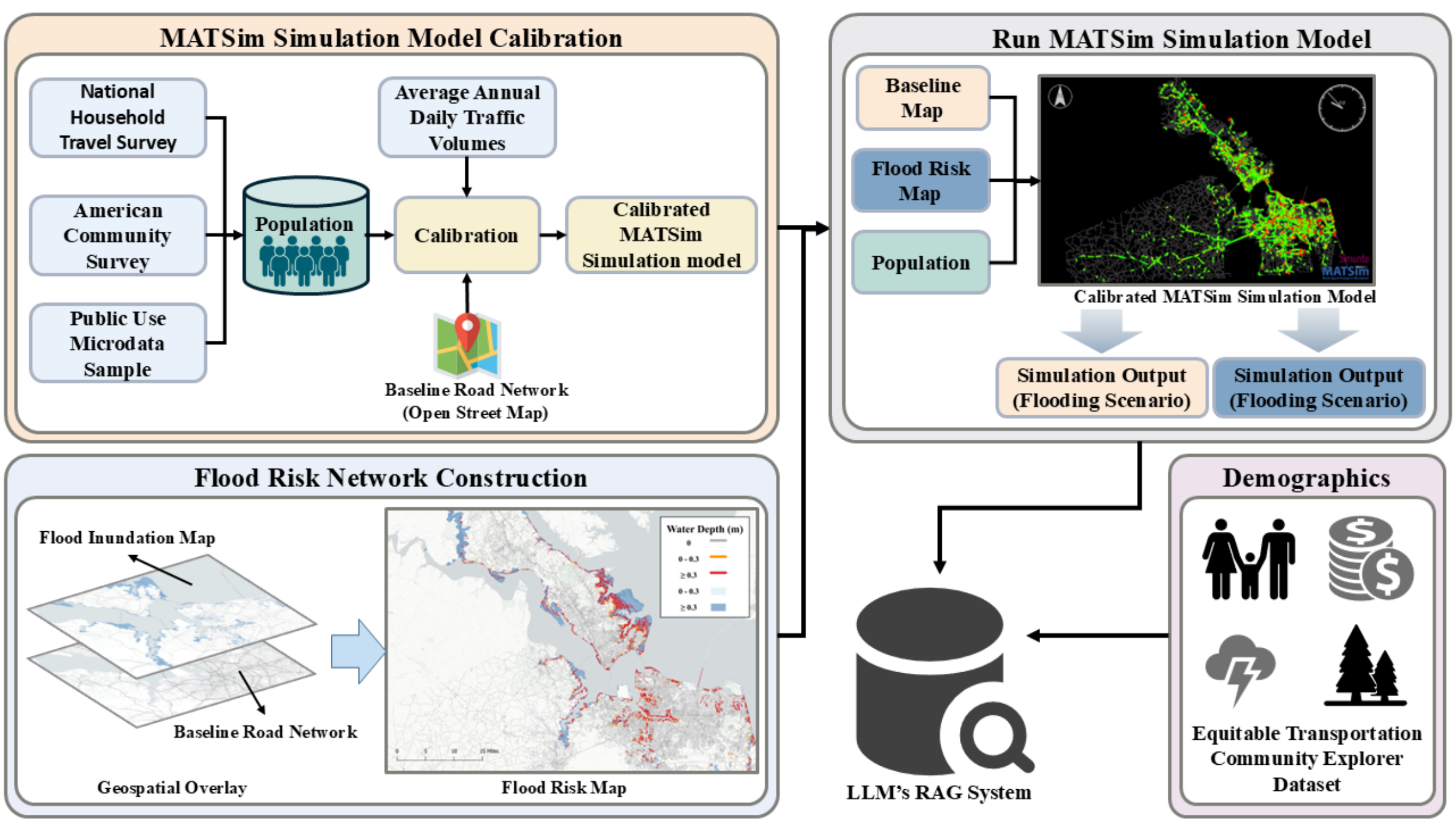

**Figure 1: The overview of data processing**

The baseline (non-flooded) road network for the Hampton Roads region was extracted from OpenStreetMap. The flood risk network was derived by overlaying the baseline road network with a flood inundation map. Link-level free-flow speed and capacity were then updated, thereby incorporating flood impacts into the flood risk network. The flood risk network served dual roles: as the modified road network input for the flood-scenario MATSim simulation, and as a structured geospatial dataset indexed within the LLM's RAG knowledge base to support link-level flood impact queries.

This study utilized an agent-based traffic simulation framework, MATSim, to run region-wide simulations on the prepared networks and produce traffic simulation information. The primary input data for the MATSim simulation model were road networks and a synthesized population which was derived from multiple datasets, including tract-level marginal distributions from the American Community Survey, household-level microdata from the Public Use Microdata Sample, and travel diary records from the National Household Travel Survey. The simulation model was calibrated using Average Annual Daily Traffic Volumes collected by the Virginia Department of Transportation.

Comparative simulation outputs under baseline and flooding scenarios were produced by running the calibrated model on both the baseline and the flood risk networks. The MATSim simulation outputs covered multiple dimensions, such as link-level traffic states, trip-level records, and person-level utilities. This study processed the simulation outputs and derived a set of tract-level indicators for resilience assessment. Specifically, vehicle miles traveled, average speed, average delay (volume-weighted excess travel time over free-flow at the link level), and trip counts characterized the flood-induced impacts on the regional transportation system operations. Mobility loss quantifies the economic impact of flooding by calculating the monetized reduction in travel utility between the baseline and flood scenarios. These indicators are summarized in Table 1. As evidenced by the mean values, the network operational efficiency and residents' travel utility showed an overall reduction in the flooding scenario relative to the baseline scenario.

**Table 1 Summary of traffic simulation indicators**

| Variable | Unit | Mean |
|---|---|---|
| Vehicle Miles Traveled Baseline | veh·mile | 84880.65 |
| Vehicle Miles Traveled Flood | veh·mile | 66377.34 |
| Average Speed Baseline | mph | 24.94 |
| Average Speed Flood | mph | 21.87 |
| Average Delay Baseline | min | 0.07 |
| Average Delay Flood | min | 0.08 |
| Trip Counts Baseline | trips | 3401 |
| Trip Counts Flood | trips | 3036 |
| Mobility Loss | USD | 18758 |
| Average Mobility Loss | USD/person | 98 |

Demographic and contextual information was sourced from the U.S. Department of Transportation's Equitable Transportation Community Explorer dataset. The ETCE dataset is a census-tract-based compilation integrating multiple federal data sources, and it characterizes cumulative disadvantage across domains such as transportation insecurity, socioeconomic vulnerability, environmental burden, and disaster risk. A subset of indicators was selected to provide tract-level social and environmental context for the LLM's RAG system, and Table 2 lists the selected variables, including their field names and corresponding descriptions.

Table 2 Descriptions of selected demographic variables

| Field Name | Unit | Description |
|---|---|---|
| trctfp | - | Census 2020 Tract FIPS code |
| cntyid | - | County ID |
| cnty | - | County name |
| ppr80h | % | Percent of houses built before 1980 |
| ppvrty | % | Percent of population with income below 200% of poverty level |
| pnmply | % | Percent of people aged 16+ unemployed |
| p65ldr | % | Percent of population aged 65 years or older |
| p17yng | % | Percent of population aged 17 years or younger |
| plmng | % | Percent of population aged 5+ with limited English proficiency |
| annlls | USD/year | Estimated annualized loss due to disasters |
| pctnnd | % | Percent of tract area inundated by 0.5 sea level increase by 2100 |
| mnmp | % | Average percentage of land classified as impervious surface per tract |

## 4. Methodology

The overall architecture of TransResAI is illustrated in Figure 2. The system is organized as a compound AI pipeline comprising four functional modules—Input, Task Decomposition, Information Retrieval, and Response Generation—coordinated by a locally deployed LLM that serves as the central reasoning engine. The LLM mediates each stage of the pipeline by interpreting natural-language queries, generating executable analytical code, orchestrating multi-source retrieval, and composing interpretable responses, enabling users to conduct flood-aware resilience analyses through plain-text interaction without engaging directly with the underlying simulation, geospatial, or document infrastructure. The pipeline draws upon four locally managed data sources: MATSim flood-scenario simulation outputs (link-level performance under baseline and flood scenarios), an OpenStreetMap-derived flood risk network (road topology with elevation and flood-risk attributes), equity-focused ETCE indicators at the census-tract level, and a regional document corpus comprising planning reports, policy documents, and technical guidelines used as the RAG knowledge base. All analytical operations execute within a SafeContext sandbox that enforces five security constraints—whitelisted libraries, read-only data access, no internet access, bounded resource consumption, and process isolation—ensuring that the entire workflow remains auditable and self-contained within agency-controlled computational environments.

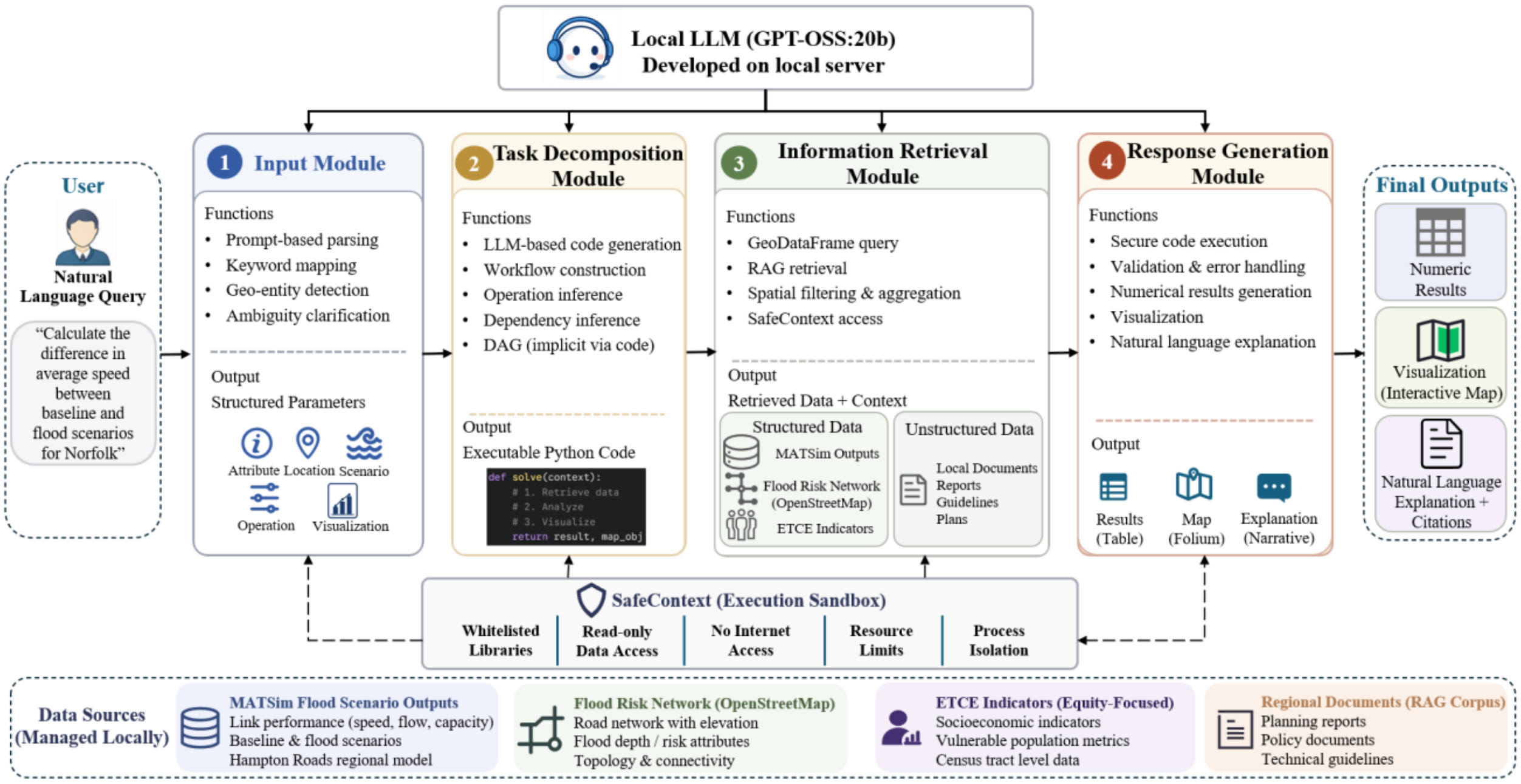

**Figure 2: LLM Reasoning Engine Architecture**

The system employs GPT-OSS:20B (Agarwal *et al.* 2025) as the base LLM, selected for its competitive performance on natural-language reasoning benchmarks while maintaining efficient inference on standard computing hardware. Structured output generation is enforced throughout the pipeline to produce deterministic responses, particularly when generating executable Python code for downstream data analysis. Upon receiving a natural-language query—for example, "Calculate the difference in average speed between baseline and flood scenarios for Norfolk"—the Input Module parses the query into five structured parameters (target attribute, geographic location, scenario type, analytical operation, and visualization specification) through prompt-based parsing, keyword mapping, geo-entity detection, and ambiguity clarification. The Task Decomposition Module then transforms these parameters into a directed acyclic workflow via LLM-based code generation, workflow construction, operation inference, and dependency inference, producing an executable function of the form def solve(ctx): whose computational graph is implicitly encoded in the generated code structure. The Information Retrieval Module accesses the integrated knowledge base through GeoDataFrame queries, RAG retrieval, and spatial filtering and aggregation, returning structured tabular results and unstructured policy passages via SafeContext-mediated access. Finally, the Response Generation Module executes the generated code within the sandbox, applies validation and error handling, and synthesizes three integrated output formats: numerical results in tabular form, interactive choropleth maps rendered with Folium, and natural-language explanations grounded with citations to retrieved policy sources. This integrated design enables practitioners without specialized GIS or simulation expertise to conduct multi-dimensional resilience assessments entirely through conversational interaction.

### 4.1. Input Module

The Input Module serves as the primary interface that transforms natural language queries into structured computational parameters through a multi-stage semantic parsing process. This module comprises four key components implemented in the GeoDataAnalyzer class: prompt-engineered parser, keyword recognition, geo-entity recognition, and ambiguity detection. The module is designed to minimize technical barriers by accepting plain English descriptions without requiring domain-specific syntax or programming knowledge.

The prompt-engineered parser leverages the LLM's natural language understanding capabilities through carefully designed system prompts following the instruction-tuning paradigm. Formally, given a user query $q \in \mathcal{Q}$, the parser generates a structured representation $\mathbf{s} = f_{\text{LLM}}\left(q, \mathcal{P}_{\text{sys}}, \mathcal{C}\right)$, where $\mathcal{P}_{\text{sys}}$ denotes the system prompt template and $\mathcal{C}$ represents the conversational context. The structured output $\mathbf{s}$ is constrained to a predefined schema containing five essential elements:

$$\mathbf{s} = \{m, \sigma, \phi, \omega, \nu\} \tag{1}$$

where $m \in \mathcal{M}$ represents the target metric from available metrics (speed, delay, vehicle miles traveled, trips, mobility loss); $\sigma \in \{$ baseline, flood $\}$ denotes the scenario type; $\phi$ represents geographic filter predicates over the geographic entity space $\mathcal{G}$; $\omega \in$ {top_k, threshold, aggregate, compare} specifies the analytical operation type; and $\nu$ indicates visualization requirements. The system translates the visualization parameter $\nu$ into a structured map specification through the _generate_default_map_spec method, which is then orchestrated alongside the codegen_system prompt to ensure the output aligns with the user's mapping intent. The code generation prompt explicitly instructs the LLM to output a single function def solve(ctx): with detailed rules for SafeContext usage, column access patterns, and error prevention. Adopting the Chain-of-Thought paradigm , the system prompts the LLM to execute explicit intermediate reasoning, thereby ensuring that complex analytical workflows are logically decomposed before final code generation (Kojima *et al.* 2022).

The keyword recognition component implements domain-specific semantic mapping through predefined synonym dictionaries. The system maintains three core mapping structures: ATTR_SYNS maps attribute synonyms to canonical forms (e.g., "velocity" and "travel speed" map to "average speed"); SCEN_SYNS handles scenario terminology (e.g., "flooded" and "storm" map to "flood"). The _token_to_col method implements a hierarchical resolution strategy consisting of exact column name matching, case-insensitive matching, synonym dictionary lookup, and fuzzy matching based on substring containment. This multi-stage approach combines explicit dictionary mapping with embedding-based similarity using the BAAI/bge-small-en-v1.5 model.

The geo-entity recognition component automatically identifies geographic references through the _detect_filters_from_question method implementing a cascaded matching strategy with three stages. First, FIPS code extraction employs pattern matching for 10 to 15 digit census tract identifiers using regex \ b(\ d{10,15}) \ b, validated against existing values in the "Census 2020 Tract FIPS code" column. Second, county name detection is performed through iterative matching of capitalized word sequences against entries in the "County name" column using exact string matching. Third, text column fuzzy matching operates on filterable text columns (those with fewer than 2,000 unique values) through token-based substring matching with minimum length threshold of 3 to 4 characters. For ambiguous references where multiple potential matches exist, the system prompts users for clarification through the conversational interface rather than making assumptions.

The ambiguity detection mechanism is implemented through intent detection methods (_detect_viz_intent, _detect_analysis_intent, _detect_rag_intent ) that scan for trigger keywords. When essential information is missing from the query, the system generates targeted clarification questions through template-based natural language generation. For instance, if a user asks "What is the speed in Norfolk?" without specifying scenario type, the system responds by inquiring whether the user would like to see the average speed under baseline conditions, flood conditions, or compare both scenarios. This interactive disambiguation process ensures the system operates on well-defined parameters while maintaining conversational fluency. The output is a structured parameter dictionary providing a standardized interface between the natural language frontend and the computational backend.

## 4.2. Task Decomposition Module

The Task Decomposition Module receives structured parameters s from the Input Module and constructs a formal computational workflow $\mathcal{W} = \{t_1, t_2, \ldots, t_n\}$ consisting of well-defined subtasks $t_i$, where each subtask represents an atomic operation with explicit input-output specifications. This module leverages the LLM's reasoning capabilities through the llm_generate_code method to generate executable Python code tailored to specific analytical tasks without requiring pre-programmed templates for every possible scenario.

The workflow builder component generates a directed acyclic graph $\mathcal{G}_{\text{task}} = (\mathcal{V}_{\text{task}}, \mathcal{E}_{\text{task}})$ representing the sequence of operations required to answer the user's query. Each vertex $v \in \mathcal{V}_{\text{task}}$ corresponds to a discrete computational step such as data filtering, metric calculation, or aggregation. Edges $e_{ij} \in \mathcal{E}_{\text{task}}$ represent data dependencies between tasks, ensuring operations are executed in topologically sorted order through the dependency relation:

$$e_{ij} \in \mathcal{E}_{\text{task}} \Leftrightarrow \text{output}(t_i) \in \text{input}(t_j) \tag{2}$$

The system analyzes structured parameters to identify required operations through the _detect_metric_hints method, which extracts scenario type, attribute, and aggregation method from query text. For standard analytical operations, the system maintains a set of task templates $\mathcal{T} = \{T_{\text{filter}}, T_{\text{rank}}, T_{\text{compare}}, T_{\text{aggregate}}\}$ combined with LLM-based reasoning for novel query patterns. For queries requesting "top 10 areas by mobility loss in high-poverty zones," the ranking template is defined as:

$$T_{\text{rank}}(\mathbf{s}) = \begin{cases} t_{\text{filter}} \rightarrow t_{\text{retrieve}} \rightarrow t_{\text{sort}} \rightarrow t_{\text{select}} & \text{if } \phi \neq \emptyset \\ t_{\text{retrieve}} \rightarrow t_{\text{sort}} \rightarrow t_{\text{select}} & \text{if } \phi = \emptyset \end{cases} \tag{3}$$

The workflow instantiation proceeds as: $t_1$ applies poverty threshold filter via ctx.subset(), $t_2$ retrieves mobility loss values via ctx.col(), $t_3$ applies sorting or ranking operations, and $t_4$ returns the top k results.

For multi-faceted queries involving scenario comparisons such as "Compare average speed reduction between baseline and flood scenarios for areas with poverty rates above 30 percent," the Task Decomposition Module generates a workflow with explicit intermediate data dependencies:

$$t_1 : \mathcal{D}_\phi \leftarrow \text{filter}(\mathcal{D}, \text{poverty_rate} > 0.30) \quad (4)$$

$$t_2 : \mathbf{v}_{\text{base}} \leftarrow \text{retrieve}\big(\mathcal{D}_\phi, m = \text{speed}, \sigma = \text{baseline}\big) \quad (5)$$

$$t_3 : \mathbf{v}_{\text{flood}} \leftarrow \text{retrieve}\big(\mathcal{D}_\phi, m = \text{speed}, \sigma = \text{flood}\big) \quad (6)$$

$$t_4 : \mathbf{v}_\Delta \leftarrow \text{pct}_{\text{change}\,(\mathbf{v}_{\text{flood}}, \mathbf{v}_{\text{base}})} = \frac{\mathbf{v}_{\text{flood}} - \mathbf{v}_{\text{base}}}{\mathbf{v}_{\text{base}}} \times 100 \quad (7)$$

$$t_5 : \mathcal{S} \leftarrow \text{aggregate}(\mathbf{v}_\Delta) = \{\mu_\Delta, \tilde{\mu}_\Delta, \sigma_\Delta\} \quad (8)$$

$$t_6 : \mathcal{R} \leftarrow \text{report}_{\text{gen}(\mathcal{S}, \mathbf{v}_\Delta)} \quad (9)$$

where $\mu_\Delta$, $\tilde{\mu}_\Delta$, and $\sigma_\Delta$ denote the mean, median, and standard deviation of percentage changes, respectively.

The operation inference component determines appropriate computational operations through keyword analysis embedded in the LLM prompt. Terms like "change," "difference," or "impact" trigger scenario comparison operations. When percentage-related terms ("percent," "rate," "%") appear in the query, the system calculates percentage change; when absolute difference terms ("absolute," "difference in") are present, absolute difference is computed; otherwise percentage change serves as the default. Ranking operations are inferred from terms like "highest," "lowest," "top," or "most," with sort direction determined by semantic polarity mapping where terms such as "highest" and "most" map to descending order while "lowest" and "least" map to ascending order. The step sequencing is implicitly handled through the generated Python code structure, where variable dependencies ensure correct execution order. For queries with multiple independent analysis branches, the code generator structures independent retrieval calls that can potentially execute concurrently. The output is executable Python code conforming to the signature def solve(ctx): -> Dict[str, Union[int, float, str]], which is validated and executed by the Response Generation Module.

### 4.3. Information Retrieval Module

The Information Retrieval Module executes computational workflows by accessing multisource data repositories $\mathcal{D} = \{\mathcal{D}_{\text{struct}}, \mathcal{D}_{\text{unstruct}}\}$ and performing specified operations. This module implements a hybrid architecture combining structured data query capabilities for quantitative datasets with Retrieval-Augmented Generation (RAG) for unstructured policy documents.

The structured data query component operates on a GeoDataFrame $\mathcal{D}_{\text{struct}}$ loaded from 'Hampton_Roads_add_columns.gpkg' containing integrated traffic simulation outputs, flood risk networks, and demographic indicators with spatial geometry information. Formally, $\mathcal{D}_{\text{struct}} = \{(\mathbf{x}_i, g_i, \mathbf{a}_i)\}_{i=1}^{N}$ where $\mathbf{x}_i \in \mathbb{R}^d$ represents the $d$-dimensional attribute vector for geographic unit $i$, $g_i \in \mathcal{G}$ denotes its geometric representation, and $\mathbf{a}_i$ contains auxiliary metadata. The '_apply_filter' method translates filter conditions into pandas operations supporting logical operators (AND, OR, NOT) and comparison operators (==,>=,<=,>,<,!=). Compound filters are evaluated through element-wise logical operations:

$$\phi_{\text{compound}} = \bigwedge_{i=1}^{k_\wedge} p_i \vee \bigvee_{j=1}^{k_\vee} p'_j \quad (10)$$

For example, a filter specifying "county equals Norfolk AND poverty rate greater than 30 percent" is compiled to:

$$\phi = \mathbb{I}[\mathbf{a}_{\text{county}} = \text{'Norfolk'}] \odot \mathbb{I}[\mathbf{x}_{\text{poverty}} > 0.30] \quad (11)$$

where $\odot$ denotes element-wise logical AND. For spatial filters involving geometric operations, the system leverages GeoPandas spatial predicates. The GeoDataFrame maintains coordinate reference system WGS84 (EPSG :4326) through the _ensure_wgs84 method, enabling geographic computations such as selecting census tracts within radius $r$ of point $p_0$:

$$\mathcal{D}_{r,p_0} = \{(\mathbf{x}_i, g_i, \mathbf{a}_i) \colon \text{distance}\ (g_i, p_0) \leq r\} \quad (12)$$

where distance$(g_i, p_0)$ computes geodesic distance using the Haversine formula. The primary data sources integrated into $\mathcal{D}_{\text{struct}}$ encompass three categories. Traffic Simulation Data ( $\mathcal{D}_{\text{traffic}}$ ) comprises link-level and tract-level metrics derived from MATSim agent-based simulations, covering 459 census tracts across 2 scenarios (baseline and flood) with 5 metrics (vehicle miles traveled, average speed, average delay, trip count, mobility loss), totaling 4,590 data points. The mobility loss metric $L_i$ for tract $i$ quantifies the monetized utility reduction based on individual travel utility differences between scenarios:

$$L_i = \sum_{p \in \mathcal{P}_i} [U_p^{\text{base}} - U_p^{\text{flood}}] \quad (13)$$

where $\mathcal{P}_i$ denotes the population in tract $i$ and $U_p^\sigma$ represents individual $p$ 's travel utility under scenario $\sigma$ derived from the MATSim scoring function. Flood Risk Network Data ( $\mathcal{D}_{\text{flood}}$ ) encompasses the full road network of 581,298 links, of which 30,570 links are identified as flood-affected and assigned capacity reduction factor $\rho_{\text{cap}} \in [0,1]$ and speed reduction factor $\rho_{\text{speed}} \in [0,1]$. For flooded links, effective capacity and free-flow speed are computed as $C_{\text{eff}} = \rho_{\text{cap}} \times C_{\text{base}}$ and $v_{\text{eff}} = \rho_{\text{speed}} \times v_{\text{base}}$ respectively. Demographic Data ( $\mathcal{D}_{\text{demo}}$ ) encompasses census tract-level socioeconomic and vulnerability indicators from the ETCE dataset, covering 459 tracts with 12 variables including poverty rate, elderly population percentage ( $\geqslant$ 65 years), limited English proficiency, unemployment rate, pre-1980 housing stock, estimated annualized disaster loss, and projected sea level rise inundation (percent of tract area inundated under 0.5 m SLR by 2100).

The RAG component is implemented through the _build_rag_engine method using LlamaIndex framework. Document preprocessing loads regulatory and planning documents from local directories and builds a vector index using Settings.embed_model = HuggingFaceEmbedding(model_name="BAAI/bge-small-en-

v1.5") and Settings.llm = self.llm . Policy documents from agencies including the Hampton Roads Planning District Commission (HRPDC), Virginia Department of Transportation (VDOT), Virginia Institute of Marine Science (VIMS), National Oceanic and Atmospheric Administration (NOAA), and Federal Emergency Management Agency (FEMA) are parsed and segmented into text chunks $\{c_1, c_2, \ldots, c_M\}$ with adaptive chunking where chunk size $|c_j| \in [500,1000]$ tokens is determined by semantic coherence. Each chunk $c_j$ is processed through the BAAI/bge-small-en-v1.5 embedding model to generate 768-dimensional semantic vectors:

$$\mathbf{e}_j = \mathcal{E}_{\text{emb}}(c_j), \mathbf{e}_j \in \mathbb{R}^{768} \tag{14}$$

These vectors are stored in a FAISS-indexed vector database $\mathcal{V}_{\text{DB}} = \{(\mathbf{e}_j, c_j, \mathbf{m}_j)\}_{j=1}^{M}$ where $\mathbf{m}_j$ contains metadata including document source, section title, and publication date. At query time, user questions $q_{\text{policy}}$ are embedded using the same model, and cosine similarity search identifies the top- $k$ most relevant chunks:

$$\mathcal{C}_{\text{retrieved}} = \arg\max_{\mathcal{C}'|=k} \sum_{c_j \in \mathcal{C}'} \frac{\mathbf{e}_q \cdot \mathbf{e}_j}{\|\mathbf{e}_q\| \|\mathbf{e}_j\|} \tag{15}$$

where $k \in [3,5]$ is dynamically determined based on query complexity. The retrieved passages are concatenated and injected into the LLM context as augmented knowledge. The final response is generated through RAG-enhanced prompting where the LLM synthesizes quantitative analysis results with relevant policy context.

The SafeContext Sandbox component provides a secure execution environment through a controlled abstraction layer. The SafeContext wrapper $\mathcal{S}_{\text{ctx}}$ exposes a restricted interface subject to security constraints: data immutability ensures the underlying GeoDataFrame $\mathcal{D}_{\text{raw}}$ remains unmodified across all operations; operation whitelist restricts exposed methods to $\mathcal{A}_{\text{safe}} = \{\text{col, get_name, subset}\}$; resource bounds limit memory usage to 8 GB and execution time to 30 seconds per operation. The SafeContext implements three primary methods. Column Access retrieves column data with automatic name resolution through a hierarchical matching strategy: first attempting exact matches against column names, then computing cosine similarity between the query alias and existing column names using the embedding model with a similarity threshold of 0.85 (determined empirically during development to balance precision and recall), and finally falling back to fuzzy substring matching if neither exact nor similarity-based matching succeeds. Name Resolution returns the actual column identifier using the same hierarchical matching strategy. Subset Filtering applies filter predicates and returns a new SafeContext instance containing the filtered data. This design ensures dynamically generated code cannot access unauthorized data, modify the original dataset, or execute potentially harmful operations such as file I/O or network requests. All operations are logged for auditability with timestamps and parameter traces, and resource limits prevent resource exhaustion through computationally expensive operations.

### 4.4. Response Generation Module

The Response Generation Module synthesizes information from the Information Retrieval Module into comprehensive, user-interpretable outputs through a multi-stage generation pipeline. This module transforms raw computational results $\mathcal{R}_{\text{raw}}$ into three integrated output formats: structured numerical reports $\mathcal{R}_{\text{num}}$, natural language explanations $\mathcal{R}_{\text{nl}}$, and interactive spatial visualizations $\mathcal{R}_{\text{vis}}$. The response generation function is formally defined as:

$$\mathcal{R} = \mathcal{G}(\mathcal{R}_{\text{raw}}, q, \mathcal{W}) = \{\mathcal{R}_{\text{num}}, \mathcal{R}_{\text{nl}}, \mathcal{R}_{\text{vis}}\} \tag{16}$$

where $q$ represents the original user query and $\mathcal{W}$ denotes the executed workflow providing context for interpretation.

The structured output generation component formats computational results with metric-specific precision through the _get_metric_display_name method. Numerical results are presented with appropriate precision determined by metric type: monetary values display two decimal places, percentages and ratios display one decimal place, and counts display whole numbers. For simple queries returning single scalar values, output is formatted as key-value pairs where the key provides a descriptive label and the value is the rounded numeric result.

The Python code generation component is orchestrated by llm_generate_code, which constructs prompts incorporating five essential components. Schema Documentation provides complete listing of 67 available data columns with descriptions and examples. API Specifications detail SafeContext method usage with usage examples and common error patterns to avoid, emphasizing critical constraints such as never nesting column access operations inside DataFrame indexing. Constraint Specifications encode prohibited operation patterns including file I/O, network access, and arbitrary code execution, along with output format requirements. Code Templates provide parameterized patterns for common operations including filtering, aggregation, ranking, and scenario comparison. Error Prevention Guidelines codify rules derived from analyzing failed generation attempts during system development. The code generation follows a constrained generation paradigm:

$$\mathcal{C}_{\text{gen}} = f_{\text{LLM}}\left(q, \mathcal{W}, \mathcal{S}_{\text{schema}}, \mathcal{P}_{\text{codegen}}, \mathcal{E}_{\text{code}}\right) \tag{17}$$

where generated code must conform to the signature def solve(ctx: SafeContext) -> Dict[str, Union[int, float, str]].

The automatic validation component is implemented through try-catch blocks in execute_generated_code. The execution environment limits the namespace to whitelisted libraries only (pandas, numpy, geopandas), with no file I/O or network access permitted. Security validation employs a whitelist-based approach maintaining permitted function calls and import statements. Multi-stage verification checks correctness criteria across three dimensions: syntax validation uses Python's AST parser to verify syntactic correctness; structural validation confirms presence of the required function signature through AST traversal; security validation scans for prohibited operations using pattern matching and AST analysis. Code is accepted only when all three validation checks pass.

The sandbox execution component runs validated code in an isolated environment $\mathcal{E}_{\text{sandbox}}$ with restricted capabilities. Execution is formalized as a bounded computation:

$$(\mathcal{R}_{\text{exec}}, \mathcal{E}_{\text{status}}) = \mathcal{E}_{\text{sandbox}}\left(\mathcal{C}_{\text{gen}}, \mathcal{S}_{\text{ctx}}, T_{\max}, M_{\max}\right) \quad (18)$$

where $T_{\max} = 30$ seconds is the time limit, $M_{\max} = 8$ GB is the memory limit, and $\mathcal{E}_{\text{status}} \in$ { success, timeout, memory_exceeded, runtime_error } indicates execution status.

The self-correction loop is implemented with a maximum of three retry attempts. When execution fails, error feedback is injected into the next generation attempt. Let $\mathcal{C}_{\text{gen}}^{(i)}$ denote code generated at iteration $i$. The self-correction process is formalized as:

$$\mathcal{C}_{\text{gen}}^{(i+1)} = f_{\text{LLM}}\left(q, \mathcal{W}, \mathcal{C}_{\text{gen}}^{(i)}, \mathcal{E}_{\text{error}}^{(i)}, \mathcal{P}_{\text{correct}}\right) \quad (19)$$

where $\mathcal{E}_{\text{error}}^{(i)}$ = { error_type, error_msg, traceback } captures exception information and $\mathcal{P}_{\text{correct}}$ is a correction-specific prompt template. The loop terminates when execution succeeds or the maximum number of iterations is reached. Common correction strategies are categorized by error type. Column name errors trigger fuzzy matching where the correction strategy switches to normalized Levenshtein similarity. Type errors lead to insertion of explicit type conversions before operations requiring numeric types. Empty data errors result in injection of defensive checks before element access operations. Logic errors detected through semantically invalid values (e.g., negative counts) trigger formula revision.

For visualization, the system extracts map specifications through _extract_map_spec by parsing LLM outputs for JSON blocks, or generates specifications via _generate_default_map_spec using query hints. The specification follows a predefined JSON schema that defines the visualization type (e.g., choropleth, heatmap, scatter), target metric, optional filter conditions, top-k ranking parameter, threshold value, and color palette. An example of such a specification is provided in Figure 3, illustrating how a natural-language query is translated into a structured representation for visualization.

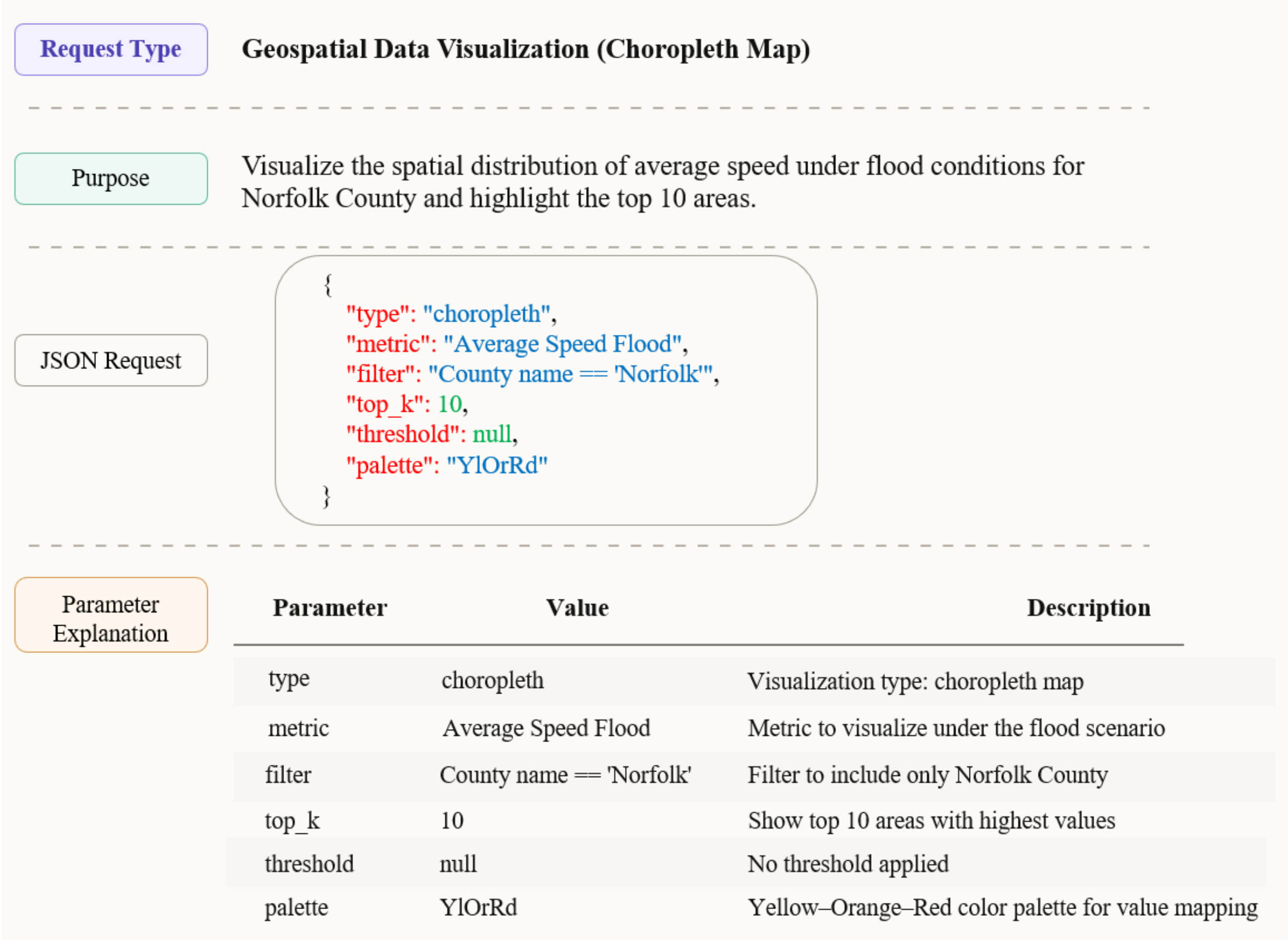

**Figure 3: Example JSON-based map specification for choropleth visualization**

Based on the generated specification, the visualization pipeline processes the request through four stages. Metric Calculation (implemented in the _compute_metric method) handles three cases: direct column access for base metrics; percentage change calculated as:

$$\mathbf{m}_{\Delta\%} = \frac{\mathbf{m}_{\text{flood}} - \mathbf{m}_{\text{base}}}{\mathbf{m}_{\text{base}}} \times 100 \tag{20}$$

and absolute difference computed as $\mathbf{m}_{\Delta} = \mathbf{m}_{\text{flood}} - \mathbf{m}_{\text{base}}$ . Color Scheme Selection (implemented in the '_resolve_cmap' method) automatically chooses sequential colormaps (Y1OrRd, Viridis) for unidirectional metrics where all values are non-negative, and diverging colormaps (RdY1Gn_r, RdBu_r) for bidirectional changes where values span both positive and negative ranges. Color mapping follows a normalized scale where for metric values with range from minimum to maximum, the color function is defined through linear interpolation mapping the normalized value to the selected colormap. Map Layer Construction (implemented in '_folium_map') creates interactive Folium maps with GeoJSON layers, using '_json_safe_gdf' for serialization with 5 decimal precision. Each geographic unit is rendered with fill color determined by its metric value, border color fixed at #555555, border weight ranging from 0.3 to 0.5 pixels based on zoom level, and fill opacity between 0.6 and 0.9 determined by data quality. Tooltip Assembly (implemented in ‘_detect_related_columns’) identifies contextually relevant variables through keyword matching. Each tooltip integrates core geographic identifiers with five auxiliary demographic variables, dynamically selected via TF-IDF cosine similarity between the user query and the ETCE dataset metadata to ensure contextual relevance. Relevance is computed through TF-IDF weighted keyword matching between query terms and variable descriptions. For instance, a query about "mobility loss" triggers inclusion of poverty rate, elderly population, unemployment, limited English proficiency, and disaster loss, while a query about "traffic speed" includes relevant infrastructure characteristics when available.

The final response combines all three formats through the Streamlit interface. The presentation template begins with introductory context explaining the analysis scope, followed by numerical results displayed via st.json(), interpretive text explaining the findings via st.markdown(), interactive visualizations rendered via components.html() at 600 pixel height, and concluding insights offering actionable recommendations. Natural language generation leverages the LLM with templates incorporating analysis results and relevant policy context from RAG. This multi-modal output design accommodates diverse user information needs and decision-making contexts.

The system is deployed on a workstation with hardware and software specifications detailed in Table 3.

**Table 3 System Implementation Specifications**

| | Component | Specification | Performance Metric |
|---|---|---|---|
| Hardware | CPU | Intel Core i9-9960X @ 3.10GHz (16 cores, 32 threads) | — |
| | GPU | Dual NVIDIA Quadro RTX 5000 (16GB VRAM each) | Tensor parallelism enabled |
| | Memory | 64GB DDR4 | Peak utilization: 6.2GB |
| Models | Base LLM | GPT-OSS:20B | 1.8s inference latency; 42.3 tokens/sec throughput |
| | Frontend Framework | Streamlit with components.html map embedding | 5 concurrent sessions supported |
| | Embedding Model | BAAI/bge-small-en-v1.5 | 768-dimensional vectors |
| | Vector Index | LlamaIndex VectorStoreIndex with FAISS backend | <100ms retrieval; persisted to disk |
| | Map Rendering | Folium + branca colormap; embedded as HTML | 600×900px; GeoJSON truncated at 5,000 records |
| Data Assets | Traffic Simulation | 459 census tracts × 2 scenarios × 10 metrics | MATSim agent-based outputs |
| | Road Network | 581298 links with flood impact attributes | OpenStreetMap + inundation model |
| | Demographics | 459 tracts × 12 vulnerability indicators | ETCE dataset (EPSG:4326 reprojection) |
| | Policy Documents | 8 documents (3.2M tokens); 6,341 index chunks | HRPDC, VDOT, VIMS, NOAA, FEMA sources |

## 5. Illustrative Examples of the User–TransResAI Interaction

TransResAI is designed to support transportation resilience analysis through natural-language interaction over heterogeneous structured and unstructured datasets. To better reflect the practical usage workflow, we organize the illustrative examples into three representative functional outputs: Analytical Answer Generation, Visualization and Spatial Plotting, and Policy-Oriented RAG Responses

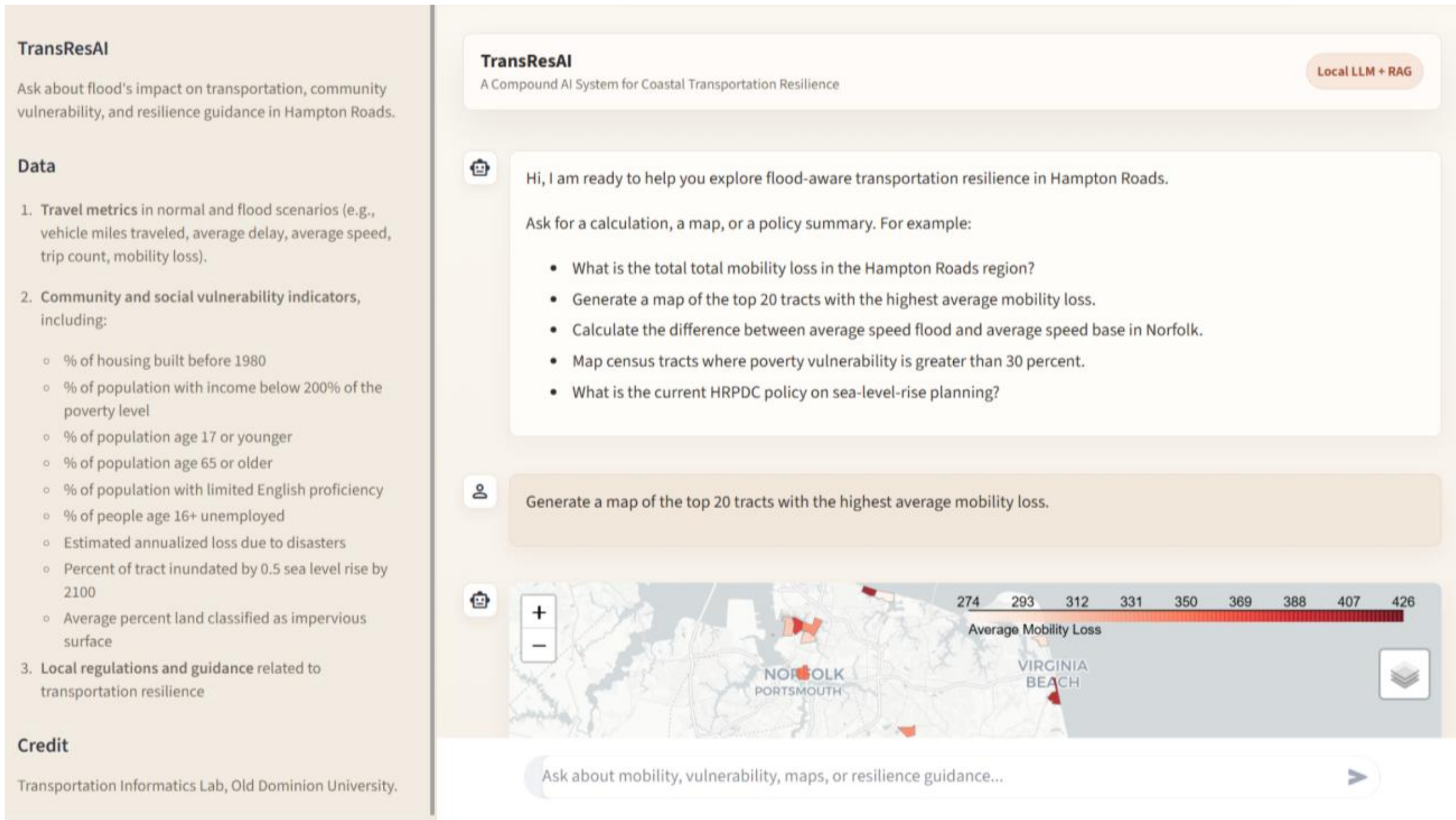

**Figure 4: TransResAI Front-End Conversational Interface**

Figure 4 illustrates the Streamlit-based conversational interface of TransResAI, including supported query categories and example analytical prompts. The system interface provides users with structured guidance regarding supported analytical dimensions, including travel performance metrics, community vulnerability indicators, and regulatory or planning documents (RAG-enabled). The example queries displayed in the interface reflect the three major interaction paradigms supported by TransResAI.

### 5.1. Analytical Answer Generation

The most fundamental interaction mode in TransResAI involves structured analytical queries that produce numerical or comparative outputs. In this mode, users pose questions related to travel performance metrics—such as average speed, delay, vehicle miles traveled, and mobility loss—under baseline and flood scenarios, as well as queries that intersect transportation impacts with demographic vulnerability indicators. Upon receiving a query, TransResAI executes a multi-stage pipeline encompassing automatic schema resolution, spatial filtering, scenario-based aggregation and comparison, and secure code execution within the SafeContext environment. The following case studies illustrate four representative interaction patterns of increasing analytical complexity, each demonstrating a distinct dimension of TransResAI's analytical capabilities.

The first case demonstrates TransResAI's ability to handle a single-area, single-metric, cross-scenario comparison—the most elemental form of resilience query. When presented with the query "Calculate the difference in average speed between the baseline and flood scenarios for Norfolk," the system automatically resolved the geographic reference to the Norfolk county filter, identified the canonical column names for baseline and flood average speed, and generated a Python solve() function to compute the mean speed difference across all relevant census tracts (see Figure 5). The calculated result indicates that Norfolk's average speed drops by approximately 3.07 mph from the baseline to the flood scenario, quantifying the direct operational impact of flooding on network mobility in this coastal city. This case demonstrates how TransResAI automates the transition from natural-language instructions to spatial analytics, providing planners with actionable numerical assessments without manual data processing.

Calculate the difference in average speed between the baseline and flood scenarios for Norfolk.

**Generated code:**

```
def solve(ctx):
    try:
        sub = ctx.subset('County name', 'Norfolk')
    except Exception:
        sub = ctx
    if sub.gdf.empty:
        return {'average_speed_difference': 0.0}
    avg_base = sub.col('Average Speed Base').mean()
    avg_flood = sub.col('Average Speed Flood').mean()
    if pd.isna(avg_base):
        avg_base = 0.0
    if pd.isna(avg_flood):
        avg_flood = 0.0
    diff = avg_base - avg_flood
    return {'average_speed_difference': diff}
```

**Calculation:**
Average speed difference (baseline – flood) = 3.0683783949151824 (rounded ≈ 3.07).
**Answer:**
Even though the regulatory guidance notes that no traffic speed data were provided, the analysis indicates that Norfolk's average speed drops by about 3.07 mph from the baseline to the flood scenario.

**Figure 5: TransResAI response to a single-area, cross-scenario speed comparison query for Norfolk**

The second case escalates the analytical complexity by introducing a spatial ranking task over a subset of the flood-scenario data. In response to the query "Which top 5 census tracts in Virginia Beach have the highest mobility loss under the flood scenario?" (see Figure 6), TransResAI dynamically resolved three distinct schema elements—Census 2020 Tract FIPS code, Total Mobility Loss, and County name—and applied a county-level spatial filter before executing a descending ranking operation via the nlargest() method. The system identified five census tracts (FIPS codes: 51810041803, 51810043600, 51810046213, 51810043800, and 51810045200) as the highest-priority areas in terms of flood-induced mobility loss. This ranking capability is particularly valuable for transportation resilience management, as it allows planners and emergency coordinators to rapidly pinpoint the geographic concentrations of travel utility loss and prioritize resource allocation accordingly, rather than treating all affected areas uniformly.

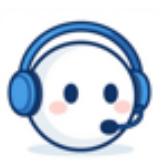

**Figure 6: TransResAI response to a spatial ranking query for top-5 high-mobility-loss census tracts in Virginia Beach**

The third case demonstrates TransResAI's capability for multi-area aggregation and cross-jurisdictional comparison. Given the query "Compare the average delay between baseline and flood scenarios across all counties. Which county experiences the greatest increase in delay?" (see Figure 7), TransResAI constructed a directed acyclic graph workflow comprising county-level groupby aggregation, flood-minus-baseline difference calculation, and index-based maximum extraction. The analysis revealed that the city of Hampton experiences the greatest flood-induced delay increase, with an average rise of approximately 0.045 minutes per vehicle (a 4.5% increase relative to the baseline). This cross-county comparison capability is essential for regional resilience planning, as it enables transportation agencies to identify which jurisdictions bear disproportionate operational burdens under flooding conditions and to calibrate mitigation strategies at the appropriate geographic scale.

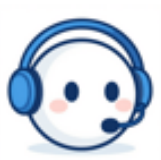

**Figure 7: TransResAI response to a cross-county delay comparison query identifying Hampton as the most impacted county**

The fourth case presents the most analytically sophisticated interaction type, combining demographic vulnerability filtering with transportation impact thresholding. Upon receiving the query "Identify census tracts with poverty rate above 30% that also experience more than 20% reduction in vehicle miles traveled under flood conditions" (see Figure 8), TransResAI simultaneously resolved four schema elements spanning both the demographic (Percent of population with income below 200% of poverty level) and simulation (Vehicle Miles Traveled under baseline and flood scenarios) datasets. The system constructed a compound Boolean mask enforcing three simultaneous conditions—poverty rate exceeding 30%, positive baseline vehicle miles traveled, and vehicle miles traveled reduction exceeding 20%—and returned the count of qualifying census tracts. This intersectional analysis capability directly addresses environmental justice concerns in transportation resilience planning, as it identifies communities that are simultaneously economically vulnerable and disproportionately exposed to flood-induced mobility loss, providing a critical input for equitable resource prioritization in disaster preparedness and post-flood recovery planning.

Identify census tracts with poverty rate above 30% that also experience more than 20% reduction in vehicle miles traveled under flood conditions.

**Generated code:**
```
def solve(ctx):
    import numpy as np
    df = ctx.gdf
    poverty_col = ctx.get_name('Percent of population with Income below 200% of poverty level')
    vmt_base_col = ctx.get_name('Vehicle Miles Traveled Base')
    vmt_flood_col = ctx.get_name('Vehicle Miles Traveled Flood')
    if df.empty:
        return {'tract_count': 0}
    poverty_mask = df[poverty_col] > 30
    base_nonzero_mask = df[vmt_base_col] > 0
    reduction_mask = (
        (df[vmt_base_col] - df[vmt_flood_col]) / df[vmt_base_col] > 0.20
    )
    combined_mask = poverty_mask & base_nonzero_mask & reduction_mask
    filtered_df = df[combined_mask]
    tract_count = len(filtered_df) if not filtered_df.empty else 0
    return {'tract_count': int(tract_count)}
```
**Calculation:**
By filtering the dataset for tracts where 'Poverty_Rate' > 0.30 and Vehicle Miles Traveled_Loss' > 0.20, the system identifies 18 unique census tracts.
**Answer:**
Based on the integrated equity-transportation analysis, 18 census tracts are identified as high-priority zones that meet both the high-poverty and significant mobility-loss criteria

**Figure 8: TransResAI response to a compound demographic-transportation query identifying high-poverty, high-vehicle miles traveled-loss census tracts**

The four cases presented above collectively illustrate TransResAI's capacity to deliver simulation-grounded, quantitatively traceable resilience assessments through natural-language interaction—addressing a critical gap in existing transportation resilience tools, which typically require specialized expertise in GIS software, simulation platforms, or programming to extract comparable insights. By integrating automatic schema resolution, spatial filtering, and secure code execution into a unified dialogue-based workflow, TransResAI enables front-line transportation planners and emergency managers to conduct sophisticated, multi-dimensional resilience analyses without requiring any direct engagement with the underlying data infrastructure.

## 5.2. Visualization and Spatial Plotting

Beyond numerical outputs, TransResAI provides an integrated spatial visualization capability that transforms analytical results into interactive choropleth maps, enabling users to perceive the geographic distribution of flood impacts and community vulnerability at a glance. When a visualization request is detected, the system automatically constructs a structured map specification in JSON format—defining the metric, spatial filter, color palette, and tooltip attributes—and passes it to the Folium-based rendering pipeline to produce an interactive HTML map embedded directly within the conversational interface. The following three cases demonstrate this visualization capability across different geographic scopes, metrics, and data domains, progressing from a single-city transportation performance indicator to a region-wide mobility impact surface and finally to a city-level long-term disaster risk assessment.

The first visualization case focuses on the intra-city spatial heterogeneity of flood-induced traffic degradation within Norfolk. In response to the query "Show the spatial distribution of average speed under the flood scenario for all census tracts in Norfolk," TransResAI applied a county-level spatial filter to isolate Norfolk's 80 census tracts, resolved the Average Speed Flood field, and rendered an interactive choropleth map using the Viridis sequential color palette (see Figure 9). The map reveals pronounced spatial variation in flood-affected travel speeds across Norfolk's census tracts; while the aggregate drop aligns with the 3.07 mph mean, localized reductions in waterfront tracts are significantly more severe, as visually indicated by the Viridis colormap. The most severely impacted tracts register speeds as low as 7.94 mph, while the least affected tracts retain speeds up to 42.74 mph. Spatially, the lowest-speed tracts are concentrated along the Elizabeth River waterfront, the Downtown area, and the Lafayette River corridor, which align with zones of high predicted inundation depth as modeled in the flood risk network. Tracts in the urban interior, particularly those encompassing elevated highway corridors such as segments of I-64 and I-264, maintain comparatively higher speeds, reflecting their relative insulation from direct flood exposure. The interactive tooltip displays the precise Average Speed Flood value for each selected tract, enabling transportation managers to identify specific neighborhoods requiring prioritized traffic management interventions without any manual GIS processing.

Show the spatial distribution of average speed under the flood scenario for all census tracts in Norfolk
Generated code:
def solve(ctx):
filtered_ctx = ctx.subset('County name', '==', 'Norfolk')
if filtered_ctx.gdf.empty:
return {'count_tracts': 0, 'mean_speed_flood': None}
count_tracts = filtered_ctx.gdf.shape[0]
mean_speed_flood = filtered_ctx.col('Average Speed Flood').mean()
return {'count_tracts': int(count_tracts),'mean_speed_flood': float(mean_speed_flood)}
Mapping process:
<MAP>{
"type": "choropleth",
"metric": "average_speed_flood",
"filter": "County name == 'Norfolk'",
"top_k": null,
"threshold": null,
"palette": "viridis"
}</MAP>
7 10 13 16 19 22 25 28 31 34
Average Speed Flood
County name: Norfolk
Average Speed Flood: 10.546
CHURCHLAND
NORFOLK
PORTSMOUTH

**Figure 9: Choropleth map of average speed under the flood scenario across all census tracts in Norfolk**

The second case expands the geographic scope to the entire Hampton Roads region, visualizing the spatial distribution of flood-induced mobility loss across all 459 census tracts. Upon receiving the query "Show a map of Total Mobility Loss under flood conditions for all census tracts in the Hampton Roads region," TransResAI rendered a region-wide choropleth mapping the Total Mobility Loss field across the full study area (see Figure 10). The aggregate regional mobility loss under flood conditions reaches 8,609,936 USD, with substantial spatial concentration: the standard deviation of tract-level losses (21,779 USD) exceeds the regional mean (18,758 USD), indicating that flood impacts are not uniformly distributed but are instead concentrated in a relatively small number of high-exposure tracts. Virginia Beach emerges as the most severely affected jurisdiction, contributing approximately 2.45 million USD in total mobility loss, with the single highest-loss tract (FIPS 51810041803) recording 187,031 USD. Norfolk and Chesapeake follow with approximately 1.66 million and 1.27 million USD respectively, while inland jurisdictions such as Williamsburg and Franklin register minimal or near-zero losses, reflecting the attenuation of flood impacts with distance from coastal waterways. Notably, the interactive tooltip simultaneously surfaces both Total Mobility Loss and Estimated Annualized Loss Due to Disasters for each selected tract—demonstrating TransResAI's capacity for context-aware visualization that automatically presents complementary risk indicators without requiring explicit user instruction.

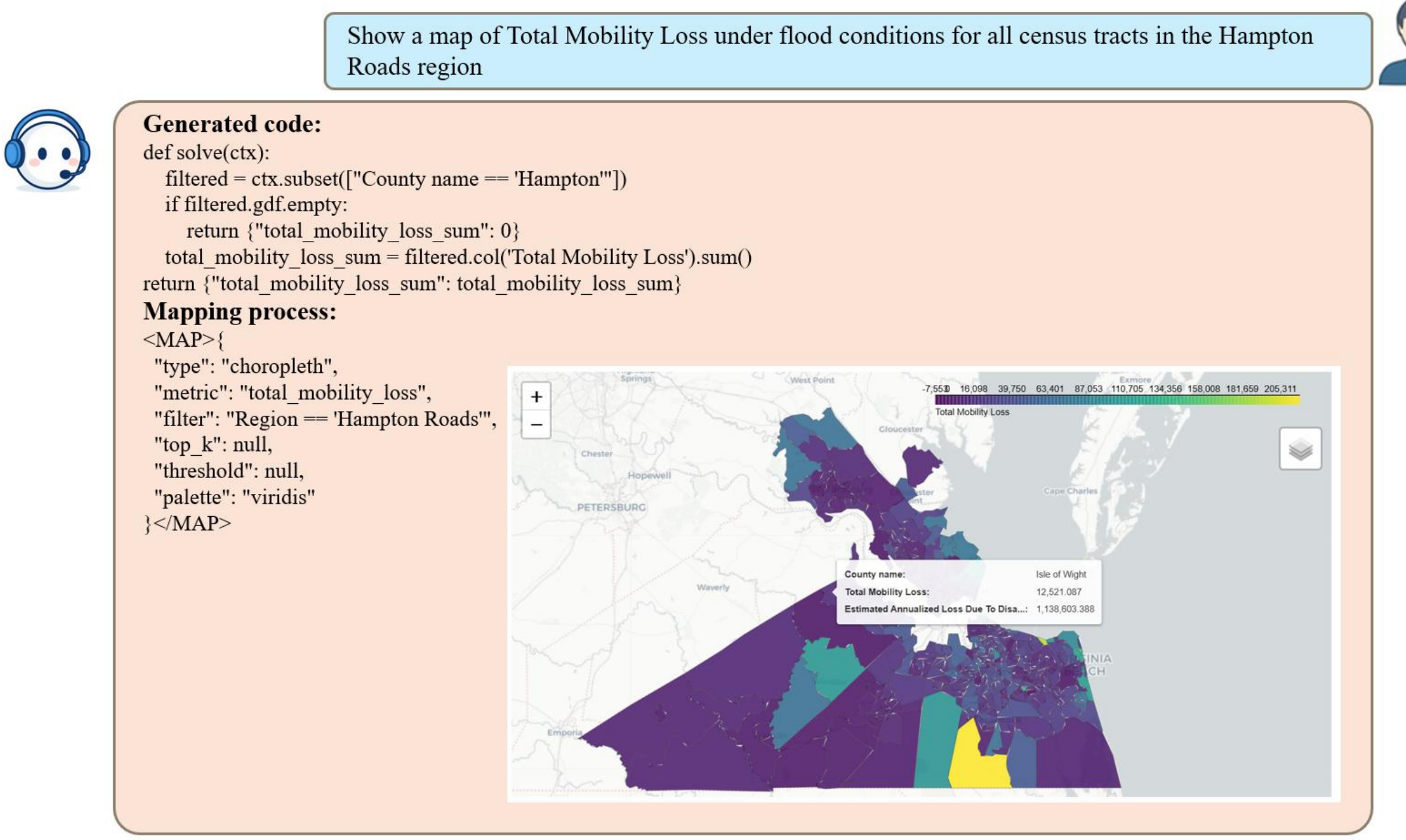

**Figure 10: Choropleth map of Total Mobility Loss under flood conditions across all census tracts in the Hampton Roads region**

The third visualization case shifts from transportation simulation outputs to the long-term disaster risk domain, illustrating TransResAI's capacity to visualize chronic hazard exposure indicators derived from the ETCE dataset. Given the query "Show the spatial distribution of estimated annualized loss due to disasters for all census tracts in Virginia Beach," the system filtered the 111 Virginia Beach census tracts and rendered a choropleth of the Estimated Annualized Loss Due to Disasters field (see Figure 11). The analysis reveals extreme spatial concentration of disaster risk: tract-level annualized losses range from a minimum of 40,417 USD to a maximum of 2,170,010 USD, with a regional mean of approximately 305,500 USD. The highest-risk tract (FIPS 51810045407) records an annualized loss more than seven times the city average, and the top five highest-loss tracts—all concentrated in the central-eastern and southern coastal portions of Virginia Beach—collectively account for a disproportionate share of the city's total chronic disaster exposure. Inland western tracts exhibit significantly lower loss values, suggesting that higher topographic elevation provides a natural buffer against flood-related economic impacts in these areas. By rendering this long-term risk layer within the same conversational interface used for flood-scenario traffic analysis, TransResAI enables users to seamlessly juxtapose short-term operational disruption with chronic disaster exposure, thereby supporting a more comprehensive and multi-temporal framing of transportation resilience planning.

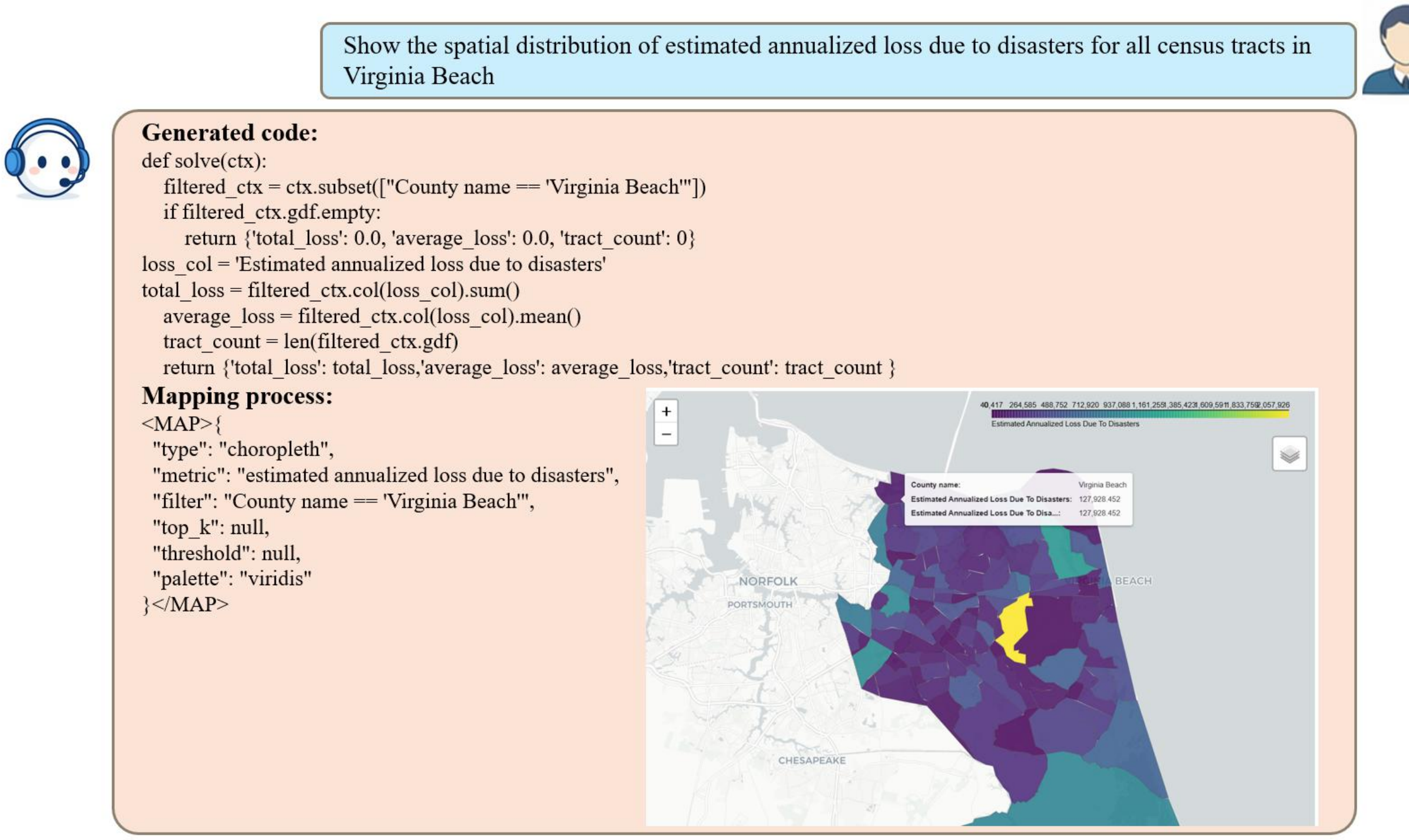

**Figure 11: Choropleth map of estimated annualized loss due to disasters across all census tracts in Virginia Beach**

The three visualization cases presented above collectively demonstrate that TransResAI's spatial plotting capability functions as an integrated geospatial reasoning layer that bridges quantitative simulation outputs and actionable spatial intelligence. Conventional GIS-based workflows for producing comparable choropleth maps would typically require users to manually join tabular simulation results with geospatial boundary files, configure color classification schemes, and manage layer rendering parameters—a process demanding both GIS proficiency and familiarity with the underlying data schema. TransResAI automates this entire pipeline through natural-language interaction, reducing the process from a multi-step technical workflow to a single conversational query while maintaining full spatial fidelity across the 459-tract study area.

## 5.3. Policy-Oriented RAG Responses

The third interaction mode supported by TransResAI enables users to query TransResAI's policy document knowledge base through natural language, retrieving and synthesizing relevant regulatory guidance, planning standards, and engineering recommendations from authoritative regional and federal sources. This capability is powered by the Retrieval-Augmented Generation (RAG) module, which indexes a curated collection of policy documents—including publications from the Hampton Roads Planning District Commission (HRPDC), the Virginia Department of Transportation (VDOT), the Virginia Institute of Marine Science (VIMS), NOAA, and FEMA—and retrieves contextually relevant passages to ground the LLM's responses in verified, source-traceable content. The following two cases illustrate how TransResAI synthesizes multi-faceted policy knowledge into structured, actionable guidance for transportation resilience planning in the Hampton Roads region.

The first case demonstrates TransResAI's ability to retrieve and synthesize a comprehensive regional planning policy from the knowledge base. In response to the query "What is the current HRPDC policy on sea-level-rise planning," TransResAI retrieved the relevant sections of the HRPDC Sea Level Rise Planning Policy (adopted October 18, 2018) and organized the response into four structured components (see Figure 12). Regarding planning benchmarks, TransResAI accurately reported the HRPDC's three-tiered scenario framework: 1.5 feet of relative sea-level rise above current mean higher high water (MHHW) for near-term planning (2018–2050), 3 feet above MHHW for medium-term planning (2050–2080), and 4.5 feet above MHHW for long-term planning (2080–2100 and beyond). On engineering and design, TransResAI conveyed that projects should employ the U.S. Army Corps of Engineers Sea Level Change Calculator to generate project-specific scenarios, followed by a benefit-cost analysis of adaptation strategies that accounts for project importance, risk tolerance, and cost. The re-evaluation component accurately noted that all scenarios and design decisions must be revisited as new information emerges from NOAA, the U.S. Army Corps of Engineers, or VIMS. Finally, TransResAI correctly identified the dual data sources underpinning the policy—NOAA's Center for Operational Oceanographic Products and Services (CO-OPS) regional projections and VIMS regression-based estimates—both of which indicate an accelerating rate of sea-level rise in the Hampton Roads area. The synthesized response concludes by characterizing the policy as a consistent, risk-based framework for localities to plan and design resilient infrastructure against projected sea-level rise, accurately reflecting the policy document's stated intent.

What is the current HRPDC policy on sea-level-rise planning

**Answer:**
The HRPDC Sea Level Rise Planning Policy (adopted October 18, 2018) calls for a staged approach to resilience:

**1. Planning Benchmarks**
Near-term (short-term) planning: 1.5 ft of relative sea-level rise above current mean higher high water (MHHW).
Medium-term planning: 3 ft above current MHHW.
Long-term planning: 4.5 ft above current MHHW.

**2. Engineering and Design**
Projects should use a tool such as the U.S. Army Corps of Engineers Sea Level Change Calculator to generate project-specific sea-level scenarios.
A benefit-cost analysis of different adaptation strategies should then determine the appropriate amount of sea-level rise to design for, taking into account project importance, risk tolerance, and cost.

**3. Re-evaluation**
Scenarios and design decisions must be revisited as new information from NOAA, the U.S. Army Corps of Engineers, or the Virginia Institute of Marine Science becomes available.

**4. Data Sources**
The policy relies on regional projections from NOAA's CO-OPS and regression-based estimates from the Virginia Institute of Marine Science (VIMS), which indicate an accelerated rise in the Hampton Roads area.

These guidelines provide a consistent, risk-based framework for localities to plan and design resilient infrastructure against projected sea-level rise.

**Figure 12: TransResAI RAG response to a query on the HRPDC sea-level-rise planning policy**

The second case narrows the policy query to the operational level, focusing specifically on engineering and design guidance. In response to the query "What are the HRPDC engineering and design recommendations for incorporating sea level rise into infrastructure projects," TransResAI retrieved the relevant engineering subsection of the same HRPDC policy document and delivered a concise response as shown in Figure 13. The system accurately identified the recommended use of the U.S. Army Corps of Engineers Sea Level Change Curve Calculator in conjunction with the 2017 NOAA sea-level-rise scenarios to generate project-specific projections, and correctly specified that a benefit-cost analysis of different adaptation options under those projected scenarios should be performed to determine the appropriate design level of relative sea-level rise, based on the project's acceptable level of risk. This focused retrieval demonstrates the RAG module's capacity for targeted subsection extraction—surfacing only the engineering-relevant content from a multi-section policy document without including extraneous planning or re-evaluation material—thereby delivering a response precisely calibrated to the specificity of the user's query.

What is the current HRPDC policy on sea-level-rise planning

**Answer:**
Use the U.S. Army Corps of Engineers Sea Level Change Curve Calculator (with the 2017 NOAA sea-level-rise scenarios) to generate project-specific sea-level-rise projections.
Perform a benefit-cost analysis of the different adaptation options under those projected scenarios to decide how much relative sea-level rise to design for, based on the project's acceptable level of risk.

**Figure 13: TransResAI RAG response to a query on HRPDC engineering and design recommendations for sea-level-rise adaptation**

The two RAG interaction cases presented above collectively illustrate TransResAI's capacity to function as an intelligent policy navigator, enabling front-line transportation managers and planners to access authoritative regulatory guidance without requiring manual review of lengthy technical documents. The contrast between the two responses is itself instructive: the first query elicited a comprehensive four-component synthesis of the full HRPDC policy framework, while the second query triggered a targeted two-point extraction focused exclusively on engineering design procedures—demonstrating that the RAG module dynamically calibrates retrieval granularity to match the scope of the user's question rather than returning fixed-length responses regardless of query specificity.

## 6. System Evaluation

To evaluate the usability, analytical effectiveness, and efficiency of TransResAI, a structured user study was conducted with domain-relevant participants performing representative resilience analysis tasks. The study assessed system performance across the three core functional categories supported by TransResAI: analytical querying over structured traffic and demographic data, spatial visualization generation, and policy-oriented retrieval-augmented generation (RAG) responses. Five participants (N = 5) were recruited with backgrounds spanning transportation engineering, urban planning, and computer science. All participants possessed prior experience with GIS-based data analysis workflows relevant to the Hampton Roads regional context, ensuring that their estimates of manual GIS completion time were grounded in practical domain knowledge and provided a meaningful efficiency comparison baseline.

Each participant independently interacted with TransResAI through natural language queries without any prior system training beyond a five-minute orientation session. Instead of following fixed scripts, each participant generated 12 spontaneous queries across three categories: Analytical (Q1–Q4), Visualization (Q5–Q8), and Policy RAG (Q9–Q12). This randomized approach ensures that the 60 recorded tasks represent diverse, realistic user intents rather than pre-determined scenarios. This open-ended questioning approach avoids the artificial task familiarity that can arise from scripted evaluations and better reflects the diversity of natural language formulations the system would encounter in real-world deployment. For each completed query, participants independently rated response quality and recorded the elapsed time. For Analytical and Visualization tasks (Q1–Q8), participants additionally provided estimates of the time they would require to perform equivalent analyses manually using conventional GIS tools, based on their own professional experience; no GIS time estimates were collected for Policy RAG tasks, as document retrieval has no direct GIS workflow equivalent.

The task categories were designed to span the full range of system capabilities, varying in complexity, output modality, and required reasoning depth. Analytical tasks required structured data queries against the integrated traffic simulation and demographic knowledge base, targeting metrics such as flood-induced mobility loss by census tract, spatial filtering by socioeconomic thresholds, scenario comparison between baseline and flood conditions, and multi-criteria ranking of vulnerable jurisdictions. Visualization tasks required the generation of interactive spatial outputs—including choropleth maps of flood-impacted road segments, equity-weighted mobility loss maps, and comparative scenario overlays—placing additional demands on the Response Generation Module's visualization pipeline and the correctness of map specification extracted from free-form queries. Policy RAG tasks involved open-ended questions requiring synthesis of regulatory guidance from regional planning agencies including HRPDC, VDOT, VIMS, NOAA, and FEMA, covering topics such as flood mitigation strategies, infrastructure investment priorities, and equity considerations.

Four metrics were recorded per task per participant. Task completion rate measured whether TransResAI produced a response that fully addressed the requested query, expressed as a proportion on a 0–1 scale. Response accuracy assessed the factual correctness and analytical soundness of the output relative to ground-truth results, rated on a 5-point Likert scale (1 = completely inaccurate, 5 = fully accurate). Answer satisfaction captured the perceived quality, clarity, and usefulness of the response from the user's perspective, also rated on a 5-point Likert scale. Task completion time measured wall-clock duration in seconds from query submission to final response display.

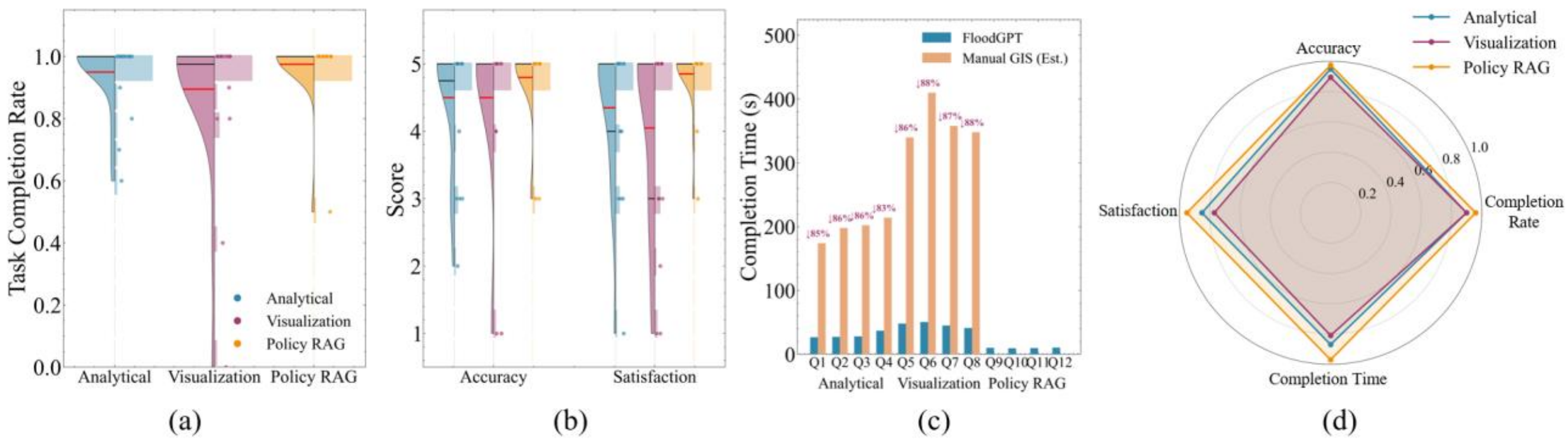

**Figure 14: Performance evaluation results of TransResAI: (a) task completion rates; (b) accuracy and satisfaction scores; (c) completion time vs. manual GIS estimation; (d) overall performance radar chart.**

As shown in Figure 14 (a), TransResAI achieved high task completion rates across all three categories. In each raincloud plot, the violin shape depicts the overall score distribution, the red horizontal line indicates the mean, the black horizontal lines denote the first and third quartiles, and the scattered dots represent individual participant responses. The Policy RAG category recorded the highest mean (0.975 ± 0.109), with the distribution tightly concentrated near 1.0, reflecting that document retrieval tasks were consistently handled with minimal failure. The Analytical category followed closely (0.950 ± 0.112), while the Visualization category exhibited a slightly lower mean (0.895 ± 0.248) with a more dispersed distribution and greater variance, attributable to occasional mismatches between user visualization intent and TransResAI's rendered output specification. Overall mean task completion exceeded 0.94 across all categories. Figure 14 (b) presents the accuracy and satisfaction score distributions using the same raincloud representation. Accuracy ratings were consistently high and clustered toward the upper end of the 1–5 scale across all three categories (Analytical: 4.50 ± 0.92; Visualization: 4.50 ± 1.20; Policy RAG: 4.80 ± 0.60), with the red mean lines positioned near 4.5–5.0, indicating that system outputs were perceived as factually reliable by domain-expert participants. The Policy RAG category achieved the highest scores on both dimensions, with user satisfaction reaching 4.85 ± 0.48 and a notably compact distribution. Satisfaction scores for the Analytical and Visualization categories were 4.35 ± 1.06 and 4.05 ± 1.40 respectively.

Figure 14 (c) compares per-question completion time between TransResAI (blue bars) and estimated manual GIS workflows (orange bars), with percentage labels annotating the time reduction achieved for each question. The contrast is visually striking: orange bars for Visualization tasks (Q5–Q8) reach heights of 340–410 seconds, while the corresponding blue bars remain below 80 seconds throughout. For Analytical tasks, TransResAI completed queries in an average of 29.7 seconds compared to an estimated manual GIS time of 197.1 seconds, representing a mean reduction of approximately 80%, with per-question speedups ranging from 83% to 86%. For Visualization tasks, the advantage was even more pronounced: TransResAI generated spatial outputs in an average of 46.1 seconds against an estimated manual GIS time of 364.0 seconds, corresponding to a mean reduction of approximately 84%, with individual speedups reaching 88% for Q6 and Q8. Policy RAG tasks (Q9–Q12) show near-zero blue bars with no GIS baseline, completed in an average of only 9.8 seconds, underscoring the efficiency of semantic vector search for large unstructured document corpora. Figure 14 (d) synthesizes normalized performance across all four dimensions as a radar chart, where each colored polygon represents one task category and a larger polygon area indicates stronger overall performance. All three polygons occupy the outer region of the chart, confirming that all categories maintained scores consistently exceeding 0.80 across accuracy, completion rate, efficiency, and satisfaction axes, with no meaningful degradation in any single dimension.

Taken together, these results demonstrate that TransResAI effectively lowers the technical barrier to complex transportation resilience analysis, delivering a mean accuracy of 4.60/5.00 and task completion rates exceeding 0.94 across all categories, while achieving efficiency gains of 80–88% over conventional manual GIS workflows. TransResAI combines task decomposition with hybrid information retrieval to automate resilience analysis. This multi-module system allows practitioners to perform expert-level assessments via conversational interaction, eliminating the need for specialized GIS or programming expertise.

## 7. Conclusion

Transportation agencies managing flood risks in coastal regions face a persistent gap between the analytical sophistication required for resilience planning and the technical accessibility available to non-specialist practitioners. TransResAI addresses this gap through a compound AI architecture that integrates LLM-driven task decomposition, secure code generation, and hybrid retrieval across simulation outputs, spatial data, and policy documents—enabling natural language access to analyses previously confined to domain experts.

Empirical validation across Hampton Roads demonstrated two principal findings. First, TransResAI reduced task completion time by 80–88% relative to conventional GIS workflows, compressing analytical operations from 197 seconds to 30 seconds without sacrificing accuracy (mean 4.60/5.00) or task completion rates (>94%). Second, the hybrid retrieval mechanism successfully synthesized quantitative traffic simulation metrics with qualitative policy context, resolving the cross-modal fragmentation that limits existing single-paradigm approaches.

These findings demonstrate that the barrier to infrastructure resilience is not a lack of data, but the technical bottleneck required to process it. TransResAI overcomes this by enabling non-expert users to perform sophisticated GIS and traffic analyses without prior programming knowledge. By reducing task completion times by 80–88%, this compound AI approach proves that the integration of task decomposition and secure code execution can democratize high-level analytics. This represents a fundamental shift: moving from a reliance on scarce technical expertise to an accessible, language-driven interface that empowers front-line planners to make informed decisions under climate uncertainty.

Future work will extend the approach across diverse coastal regions and hazard types to further assess the generalizability of the proposed framework. More broadly, this study provides a replicable foundation for expanding access to specialized infrastructure analytics, addressing a critical need that extends beyond transportation resilience to a broader range of data-intensive decision-making challenges faced by public agencies under climate uncertainty.

**Acknowledgment**

The contents of this paper present the views of the authors, who are responsible for the facts and accuracy of the data presented herein. The contents of the paper do not reflect the official views or policies of the agencies. The work was funded by the Institute for Coastal Adaptation and Resilience CARMAR program and the Transportation Informatics Lab, Department of Civil & Environmental Engineering at Old Dominion University (ODU).

**References**

Agarwal, S., Ahmad, L., Ai, J., Altman, S., Applebaum, A., Arbus, E., Arora, R.K., Bai, Y., Baker, B., Bao, H., 2025. Gpt-oss-120b & gpt-oss-20b model card. arXiv preprint arXiv:2508.10925.

Alrahamneh, A., Murshed, A.A., Alhalalmeh, A.-H., 2025. Cybersecurity in social security systems: Addressing the risks of digitalization. In: Musleh Al-Sartawi, A.M.A., Nour, A.I., Abdeljawad, I. eds. Business resilience and business innovation for sustainability: The double-edged role of artificial intelligence and other disruptive technologies. Springer Nature Switzerland, Cham, pp. 2601-2612.

Arvidsson, B., Guldåker, N., Johansson, J., 2023. A methodological approach for mapping and analysing cascading effects of flooding events. International Journal of River Basin Management 21 (4), 659-671.

Aziz, F., Wang, X., Mahmood, M.Q., Awais, M., Trenouth, B., 2024. Coastal urban flood risk management: Challenges and opportunities− a systematic review. Journal of Hydrology 645, 132271.

Bhanye, J., 2025. Flood-tech frontiers: Smart but just? A systematic review of ai-driven urban flood adaptation and associated governance challenges. Discover Global Society 3 (1), 59.

Chabot, M., Bertrand, J.-L., 2025. Adaptive flood risk management: A decision support system integrating deep learning, digital twins, and economic risk assessment. Global Environmental Change 95, 103069.

Chen, F., Chopra, S.S., 2025. Integrating socio-demographic factors for equitable resilience in networked-urban infrastructure systems. npj Urban Sustainability 5 (1), 23.

Chen, J., Ye, J., Wang, G., 2025a. From standalone llms to integrated intelligence: A survey of compound al systems. arXiv preprint arXiv:2506.04565.

Chen, K., Zhou, X., Lin, Y.-C.A., Feng, S., Shen, L., Wu, P., 2025b. A survey on privacy risks and protection in large language models. Journal of King Saud University Computer and Information Sciences 37.

Choi, M., Seo, J., Hohl, A., 2025. Agent-based travel scheduler: Decomposing od data for predicting individual travel schedules through agent-based modeling. Journal of Geographical Systems 27 (2), 257-282.

Fang, C., Chu, Y., Fu, H., Fang, Y., 2022. On the resilience assessment of complementary transportation networks under natural hazards. Transportation research part D: transport and environment 109, 103331.

Gao, Y., Xiong, Y., Gao, X., Jia, K., Pan, J., Bi, Y., Dai, Y., Sun, J., Guo, Q., Wang, M., Wang, H., 2023. Retrieval-augmented generation for large language models: A survey. ArXiv abs/2312.10997.

Gupta, K., Shrivastava, A., 2025. Zero data retention in llm-based enterprise ai assistants: A comparative study of market leading agentic ai products. ArXiv abs/2510.11558.

Gurnee, W., Tegmark, M., 2023. Language models represent space and time. ArXiv abs/2310.02207.

Habib, E.H., Miles, B., Skilton, L., Elsaadani, M., Osland, A.C., Willis, E., Miller, R., Do, T., Barnes, S.R., 2023. Anchoring tools to communities: Insights into perceptions of flood informational tools from a flood-prone community in louisiana, USA. Frontiers in Water 5, 1087076.

He, K., Carhart, N., Pregnolato, M., De Risi, R., 2026. Flood-induced traffic congestion and accessibility loss for urban road networks using agent-based simulation: The case study of bristol, uk. International Journal of Disaster Risk Reduction, 106053.

Huang, J., Cui, Y., Zhang, L., Tong, W., Shi, Y., Liu, Z., 2022. An overview of agent-based models for transport simulation and analysis. Journal of Advanced Transportation 2022 (1), 1252534.

Jana, D., Malama, S., Narasimhan, S., Taciroglu, E., 2023. Edge-based graph neural network for ranking critical road segments in a network. Plos one 18 (12), e0296045.

Ji, Y., Gao, S., Nie, Y., Majić, I., Janowicz, K., 2025. Foundation models for geospatial reasoning: Assessing the capabilities of large language models in understanding geometries and topological spatial relations. International Journal of Geographical Information Science 39, 1866 - 1903.

Jiang, X., Sengupta, R., Demmel, J., Williams, S., 2024. Large scale multi-gpu based parallel traffic simulation for accelerated traffic assignment and propagation. Transportation Research Part C: Emerging Technologies 169, 104873.

Kammouh, O., Chahrour, N., 2025. Indicator-based framework to evaluate the resilience of transport infrastructure systems. Sustainable and Resilient Infrastructure 10 (5), 536-562.

Kandogan, E., Bhutani, N., Zhang, D., Chen, R.L., Gurajada, S., Hruschka, E., Year. Orchestrating agents and data for enterprise: A blueprint architecture for compound ai. In: Proceedings of the 2025 IEEE 41st International Conference on Data Engineering Workshops (ICDEW), pp. 18-27.

Kojima, T., Gu, S.S., Reid, M., Matsuo, Y., Iwasawa, Y., 2022. Large language models are zero-shot reasoners. Advances in neural information processing systems 35, 22199-22213.

Konev, A., Sadransky, B., Rauer-Zechmeister, S., Cornel, D., Schwerdorf, I., Waser, J., Vuckovic, M., 2025. Interactive geospatial stories for flood management. ACM Transactions on Modeling and Computer Simulation 35 (4), 1-17.

Lander, M.S.F., Meguro, W., Briones, J.I., Castillo, I.R., Fletcher, C.H., 2024. Envisioning in-situ sea level rise adaptation for coastal cities. Technology|Architecture + Design 8 (2), 298-311.

Ldchn, K., Kato, T., Sano, K., 2025. Analyzing transportation network vulnerability to critical-link attacks through topology changes and traffic volume assessment. Applied Sciences 15 (8), 4099.

Lee, C.-C., Huang, L., Antolini, F., Garcia, M., Juan, A., Brody, S.D., Mostafavi, A., 2024. Predicting peak inundation depths with a physics informed machine learning model. Scientific Reports 14 (1), 14826.

Li, C., Wang, W., Solé-Ribalta, A., Borge-Holthoefer, J., Jia, B., Liu, Z., Yu, B., Gao, Z., Gao, J., 2025. Adaptive capacity for multimodal transport network resilience to extreme floods. Nature Sustainability 8 (7), 741-752.

Liang, J., Su, G., Lin, H., Wu, Y., Zhao, R., Li, Z., 2025. Reasoning rag via system 1 or system 2: A survey on reasoning agentic retrieval-augmented generation for industry challenges. ArXiv abs/2506.10408.

Lin, X., Lu, Q., Chen, L., Brilakis, I., 2024. Assessing dynamic congestion risks of flood-disrupted transportation network systems through time-variant topological analysis and traffic demand dynamics. Advanced Engineering Informatics 62, 102672.

Ling, C., Zhao, X., Lu, J., Deng, C., Zheng, C., Wang, J., Chowdhury, T., Li, Y., Cui, H., Zhang, X., 2025. Domain specialization as the key to make large language models disruptive: A comprehensive survey. ACM Computing Surveys 58 (3), 1-39.

Mohsin, M.A., Umer, M., Bilal, A., Memon, Z., Qadir, M.I., Bhattacharya, S., Rizwan, H., Gorle, A.R., Kazmi, M.Z., Mohsin, A., Rafique, M.U., He, Z., Mehta, P., Jamshed, M.A., Cioffi, J.M., 2025. On the fundamental limits of llms at scale. ArXiv abs/2511.12869.

Museru, M.L., Nazari, R., Nikoo, M.R., Karimi, M., 2025. Improving disaster resilience with causal machine learning for flood damage estimation. Science of the Total Environment 995, 180121.

Nelson, L.K., Bogeberg, M., Cullen, A., Koehn, L.E., Strawn, A., Levin, P.S., 2022. Perspectives on managing fisheries for community wellbeing in the face of climate change. Maritime Studies 21 (2), 235-254.

Nguyen, S.S., Kittur, J., Year. Review of current and potential uses of large language models in engineering. In: Proceedings of the Frontiers in Education, pp. 1649650.

Nie, T., Sun, J., Ma, W., 2025. Exploring the roles of large language models in reshaping transportation systems: A survey, framework, and roadmap. Artificial Intelligence for Transportation 1, 100003.

Nwogu, N., Victoria, M.F., Salman, H., Oyetunji, A.K., 2025. Integrating gis into flood risk management: A global south perspective on resilience, planning, and policy. Water 17 (21), 3149.

Pan, X., Dang, Y., Wang, H., Hong, D., Li, Y., Deng, H., 2022. Resilience model and recovery strategy of transportation network based on travel od-grid analysis. Reliability Engineering & System Safety 223, 108483.

Pierdicca, R., Muralikrishna, N., Tonetto, F., Ghianda, A., 2025. On the use of llms for gis-based spatial analysis. ISPRS International Journal of Geo-Information 14 (10), 401.

Prashar, N., Lakra, H.S., Shaw, R., Kaur, H., 2023. Urban flood resilience: A comprehensive review of assessment methods, tools, and techniques to manage disaster. Progress in Disaster Science 20, 100299.

Pregnolato, M., Ford, A., Wilkinson, S.M., Dawson, R.J., 2017. The impact of flooding on road transport: A depth-disruption function. Transportation Research Part D: Transport and Environment 55, 67-81.

Pu, Q., Xie, K., Yang, H., Zhai, G., 2026. A vision-and-knowledge enhanced large language model for generalizable pedestrian crossing behavior inference. arXiv preprint arXiv:2601.00694.

Rouhana, F., Jawad, D., 2025. A spatial-network approach to assessing transportation resilience in disaster-prone urban areas. ISPRS International Journal of Geo-Information 14 (7), 261.

Sharma, C., 2025. Retrieval-augmented generation: A comprehensive survey of architectures, enhancements, and robustness frontiers. ArXiv abs/2506.00054.

Singha, C., Rana, V.K., Pham, Q.B., Nguyen, D.C., Łupikasza, E., 2024. Integrating machine learning and geospatial data analysis for comprehensive flood hazard assessment. Environmental Science and Pollution Research 31 (35), 48497-48522.

Subramanian, S., 2024. Large language model-based solutions: How to deliver value with cost-effective generative ai applications John Wiley & Sons.

Truong, T.H., Lau, J.H., Qi, J., 2025. Understanding the geospatial reasoning capabilities of llms: A trajectory recovery perspective. ArXiv abs/2510.01639.

Walker, C.L., Siems-Anderson, A., Towler, E., Dugger, A., Gaydos, A., Wiener, G., 2024. Towards development of a roadway flood severity index. Transportation Research Interdisciplinary Perspectives 27, 101218.

Wampler, D., Nielson, D.F., Seddighi, A., 2025. Engineering the rag stack: A comprehensive review of the architecture and trust frameworks for retrieval-augmented generation systems. ArXiv abs/2601.05264.
Wang, P., Wu, X., 2025. Ai-driven approaches to flood risk management: Overcoming data bias and enhancing decision-making. Climate Risk Management, 100752.
Wang, S., Year. Gsi agent: Domain knowledge enhancement for large language models in green stormwater infrastructure.
Wu, F., Shen, T., Bäck, T., Chen, J., Huang, G., Jin, Y., Kuang, K., Li, M., Lu, C., Miao, J., 2025. Knowledge-empowered, collaborative, and co-evolving ai models: The post-llm roadmap. Engineering 44, 87-100.
Xu, L., Zhao, S., Lin, Q., Chen, L., Luo, Q., Wu, S., Ye, X., Feng, H., Du, Z., 2025. Evaluating large language models on geospatial tasks: A multiple geospatial task benchmarking study. International Journal of Digital Earth 18 (1), 2480268.
Yan, B., Li, K., Xu, M., Dong, Y., Zhang, Y., Ren, Z., Cheng, X., 2025. On protecting the data privacy of large language models (llms) and llm agents: A literature review. High-Confidence Computing 5 (2), 100300.
Yang, Y., Huang, H., Li, G., Han, B., Yuan, Z., Ma, H., 2025. A systematic review of resilience assessment and enhancement of urban integrated transportation networks. Journal of Transport Geography 129, 104420.
Ying, S., Li, Z., Yu, M., 2025. Beyond words: Evaluating large language models in transportation planning. Geo-spatial Information Science, 1-23.
Yuan, F., Lee, C.-C., Mobley, W., Farahmand, H., Xu, Y., Blessing, R., Dong, S., Mostafavi, A., Brody, S.D., 2023. Predicting road flooding risk with crowdsourced reports and fine-grained traffic data. Computational Urban Science 3 (1), 15.
Zaharia, M., Khattab, O., Chen, L., Davis, J.Q., Miller, H., Potts, C., Zou, J., Carbin, M., Frankle, J., Rao, N., 2024. The shift from models to compound ai systems. The berkeley artificial intelligence research blog.
Zhang, Q., Gao, S., Wei, C., Zhao, Y., Nie, Y., Chen, Z., Chen, S., Su, Y., Sun, H., 2025. Geoanalystbench: A geoai benchmark for assessing large language models for spatial analysis workflow and code generation. Transactions in GIS 29 (7), e70135.
Zhang, S., Fu, D., Liang, W., Zhang, Z., Yu, B., Cai, P., Yao, B., 2024a. Trafficgpt: Viewing, processing and interacting with traffic foundation models. Transport Policy 150, 95-105.
Zhang, Z., Wang, X., Zhang, Z., Li, H., Qin, Y., Zhu, W., Year. Llm4dyg: Can large language models solve spatial-temporal problems on dynamic graphs? In: Proceedings of the Proceedings of the 30th ACM SIGKDD conference on knowledge discovery and data mining, pp. 4350-4361.
Zhao, S., Yang, Y., Wang, Z., He, Z., Qiu, L.K., Qiu, L., 2024. Retrieval augmented generation (rag) and beyond: A comprehensive survey on how to make your llms use external data more wisely. arXiv preprint arXiv:2409.14924.
Zhou, J., Li, Z., Zhang, Y., Dong, S., 2026. Transportation infrastructure resilience and safety under climate risks: A global bibliometric and thematic analysis (1995–2025). Multimodal Transportation, 100307.
Zhu, J., Dang, P., Cao, Y., Lai, J., Guo, Y., Wang, P., Li, W., 2024. A flood knowledge-constrained large language model interactable with gis: Enhancing public risk perception of floods. International Journal of Geographical Information Science 38 (4), 603-625.
Zhuge, C., Bithell, M., Shao, C., Li, X., Gao, J., 2021. An improvement in matsim computing time for large-scale travel behaviour microsimulation. Transportation 48 (1), 193-214.